\documentclass[prodmode,acmtoit]{acmsmall}

\usepackage{url}
\usepackage{cite}
\usepackage{hypotheses}
\usepackage{caption}
\usepackage{array}
\usepackage{graphicx}
\usepackage{amssymb}
\usepackage{amsmath}
\usepackage{mathtools}
\usepackage[usenames,dvipsnames,svgnames,table]{xcolor}
\usepackage{multirow}
\usepackage{flushend}

\usepackage{wrapfig}
\usepackage{changepage}
\usepackage{mdframed}
\usepackage{ntheorem}

\definecolor{formalshade}{rgb}{0.95,0.95,1}
\newcommand{\rev}[1]{\textcolor{black}{#1}}

\theoremstyle{nonumberplain}

\begin{document}

\title{Patterns in the Chaos -- a Study of Performance Variation and Predictability in Public IaaS Clouds}	

\author{
Philipp Leitner
\affil{Department of Informatics, University of Zurich, Switzerland} 
J\"urgen Cito
\affil{Department of Informatics, University of Zurich, Switzerland}
}

\begin{abstract}
Benchmarking the performance of public cloud providers is a common research topic. Previous work has already extensively evaluated the performance of different cloud platforms for different use cases, and under different constraints and experiment setups. 
In this paper, we present a principled, large-scale literature review to collect and codify existing research regarding the predictability of performance in public Infrastructure-as-a-Service (IaaS) clouds. We formulate 15 hypotheses relating to the nature of performance variations in IaaS systems, to the factors of influence of performance variations, and how to compare different instance types. In a second step, we conduct extensive real-life experimentation on four cloud providers to empirically validate those hypotheses.
We show that there are substantial differences between providers.
Hardware heterogeneity is today less prevalent than reported in earlier research,
while multi-tenancy indeed has a dramatic impact on performance and predictability, but only for some cloud providers.
We were unable to discover a clear impact of the time of the day or the day of the week on cloud performance.
\end{abstract}




\begin{bottomstuff}
The research leading to these results has received funding from the
European Community's Seventh Framework Programme (FP7/2007-2013) under grant
agreement no. 610802 (CloudWave). We thank Joel Scheuner for valuable feedback
and help with setting up our benchmarks in Cloud Workbench.
\end{bottomstuff} 

\maketitle

\section{Introduction}
\label{sec:intro}

In an Infrastructure-as-a-Service (IaaS) cloud~\cite{armbrust:10}, computing
resources are acquired and released as a service, typically in the form of
virtual machines with attached virtual disks~\cite{buyya:09}. Cloud benchmarking, i.e., the process
of establishing and objectively comparing the performance of different providers
and configurations, is a common contemporary research topic in the cloud domain.

Previous work has already extensively
evaluated the performance of different cloud platforms for different use cases,
and under different constraints and experiment setups.
However, we observe a number of issues with the current state of research.
Firstly, most existing papers do not make all parameters that impact the
reported results, or the raw data itself, available, leading to the unsatisfying situation that similar
experiments are reported in literature with differing results. Readers are
unable to establish whether this is due to external factors (e.g., the
performance of cloud providers changing over time), unreported parameters, or
technical inaccuracies.
Secondly, current research papers generally do not compare their results to
previous work, but start from a ``clean slate''. Thirdly, it is unclear to what
extent research published, for instance, in 2010 stood the test of time, and remains
valid today. One has to consider that cloud benchmarking inherently
aims at a moving target, as cloud providers are constantly competing to improve
their service.
All in all, despite the plethora of existing data points,
it remains surprisingly difficult to extract meaningful and portable knowledge
from existing research.

The goal of our study is to identify the fundamental rules, patterns and
mechanisms underlying performance variations  of IaaS-based public cloud
systems. Specifically, we are interested in \emph{predictability} of
performance, i.e., how accurately the performance of a virtual machine acquired
from an IaaS cloud can be estimated in advance, and how stable this performance
will be.
We present a principled
literature review to collect and codify existing research on the performance
variations underlying public IaaS systems.
We formulate 15 hypotheses relating to the nature of performance
predictability, to the factors of influence of performance variations, and how
to select cloud instance types. 
Further, we systematically collect
real-life performance data from four IaaS providers (Amazon Elastic Compute Cloud, EC2\footnote{\url{http://aws.amazon.com/ec2/}},
Google Compute Engine, GCE\footnote{\url{https://cloud.google.com/compute/}}, Microsoft Azure\footnote{\url{https://azure.microsoft.com/en-us/}}, and IBM Softlayer, SL\footnote{\url{http://www.softlayer.com}}) using 5 micro and application-level benchmarks
in 2 to 8 different configurations, executed 6 times a day over a period of one
month. This led to two data sets, with a combined total of 53918 measurements.
We use this data to validate the hypotheses we formed from
existing research. We show that there are relevant differences between
providers, and illustrate that not all assumptions and results from existing
research are equally valid for all cloud providers.
For instance, we show that
hardware heterogenity, as often seen as a core property of public clouds, only exists
in Azure and a small number of EC2 instance types at the time of this writing.
Similarly, the effect of multi-tenancy on performance predictability is not equally
pronounced in all clouds.



\section{Preliminaries}
\label{sec:background}

In the IaaS model of cloud computing, virtual machines (or, as they are more
commonly referred to, \emph{instances})  can be configured along a number of
dimensions.
Most importantly, when requesting an instance, users need to
specify (1) a \emph{region} (e.g, \texttt{us-east-1}), which maps to a known physical
data center within the cloud, (2) a \emph{base image}, which defines what
software (operating system plus optional pre-installed software packages) should
initially be installed in the virtual machine, and (3) an \emph{instance type},
which defines price and available computing resources for the instance.
\begin{wraptable}{r}{0.55\textwidth}
\scriptsize
\centering
\begin{tabular}{|l|c|l|p{2.5cm}|}
\hline
\rowcolor[HTML]{EFEFEF} \textbf{Instance Type} & \textbf{CPU} & \textbf{Price} & \textbf{Common CPU} \\
  \rowcolor[HTML]{EFEFEF}   & \textbf{Cores} &  & \textbf{Models} \\
  \hline
  m1.small & 1 & 0.044\$  & Xeon E5645 \newline Xeon E5-2650  \\
  \hline
   m3.large & 2 & 0.14\$  & 2x Xeon E5-2670  \\
  \hline
	 t1.micro & $< 1$  & 0.02\$  &  Xeon E5645 \newline Xeon E5430 \newline Xeon E5507 \\
  \hline
	 i2.xlarge & 4 & 0.853\$  & 4x Xeon E5-2670 v2 \\
\hline
\end{tabular}
\caption{Example instance types, hourly prices, and reported physical CPU models as of November 2014 in EC2, US region.}
\label{tab:types}
\end{wraptable}
Instance
types are typically grouped into \emph{families} of types for comparable use
cases (e.g., general-purpose, CPU-optimized, I/O-optimized).
IaaS clouds
often serve requests for identical instance
types with differing underlying hardware (\emph{hardware heterogeneity}).
Table~\ref{tab:types} lists a subset of existing
EC2 instance types along with their hourly price as of beginning of November 2014
in the \texttt{us-east-1} region for a Linux instance.
Additionally, we list, without claim of completeness, physical CPU models
that are at the time of our study commonly used to serve a given instance type for
EC2.
In IaaS, multiple customers typically own instances running on the same physical
machine (\emph{multi-tenancy}). Most computing resources (e.g., network or disk I/O, but usually not CPU cores)  are shared among all instances on a machine. This can lead
to the ``noisy neighbour'' problem, when one instance experiences a slowdown
due to the behavior of a co-located other tenant~\cite{gkatzikis:13}.

\section{Related Work}
\label{sec:related}

Our study takes inspiration from the efforts of Li et al., who
have been the first to systematically collect and classify
research on IaaS benchmarking. Their work has led to a methodology for
evaluating providers~\cite{li:13:2}, to a catalogue of widely-used
metrics~\cite{li:12:2}, and to a conceptual model of IaaS benchmarks~\cite{li:13}.
However, while this work aptly classifies
existing research, it does not actually codify or validate the results presented
therein, which is the scope of our research.

The empirical part of
our research requires means to easily define and execute benchmarks over
different cloud systems. Previous work has proposed multiple approaches to
achieve this, including Expertus~\cite{jayasinghe:12},
CloudBench~\cite{silva:08}, CloudCrawler~\cite{cunha:13}, and Cloud
Workbench~\cite{scheuner:14,scheuner:15}. We have made use of our own framework (Cloud
Workbench) to collect data for quantitative analysis. However, arguably,
either of the other systems could have been used instead as well.

Our work is also related
to the efforts of CloudHarmony\footnote{\url{https://cloudharmony.com}}. CloudHarmony is a commerical
entity, which has been collecting numerical performance data from a large set of cloud
providers and services \rev{over a long period of time (multiple years in many cases).
Unfortunately, at the time of this writing, the CloudHarmony benchmarking data has been removed from their
website.}
The main difference between our work and the data
collected by CloudHarmony is that they are primarily striving for breadth in the
data they collect (many providers, configurations, \rev{and a longer observation period}), while we
were going for depth (less services \rev{benchmarked over a much shorter period of time},
but significantly more data per provider).

Finally, our work is evidently also closely related to previous cloud
benchmarking research, which we cover as part of
Section~\ref{sec:observations}.

\section{Hypotheses}
\label{sec:observations}

We conducted a systematic literature review~\cite{kitchenham:04}
to collect the existing state of research regarding performance predictability of public IaaS clouds.
We seeded our  review with 7 seminal publications, which we considered widely-known and representative of the field~\cite{phillips:11,schad:10,lenk:11,mao:12,iosup:11,imai:13,fittkau:12}. We also added three of our own earlier publications to the  seed~\cite{borhani:14,scheuner:14,leitner:15}. From those \rev{10} studies, we generated a candidate set of additional relevant studies by following all backward and forward scientific references. 

From the resulting set of additional candidate studies, we accepted those that were (1) specifically about benchmarking \emph{public} cloud services (we disregarded studies on private cloud systems or self-hosted data centers, as those systems are substantially different from a performance predictability point of view), (2)  which explicitly reported numerical performance data (we disregarded studies that mention benchmarking as part of their research, but do not explicitly report their results), and (3) which appeared in a trustworthy peer-reviewed scientific venue
(specifically, we excluded several publications that were published in journals on Beall's list of predatory open access journals\footnote{\url{http://scholarlyoa.com/publishers/}} -- after manual inspection the results of these papers did not appear sufficiently reliable). 
\begin{wrapfigure}{r}{0.35\textwidth}
\centering
 \includegraphics[width=\linewidth]{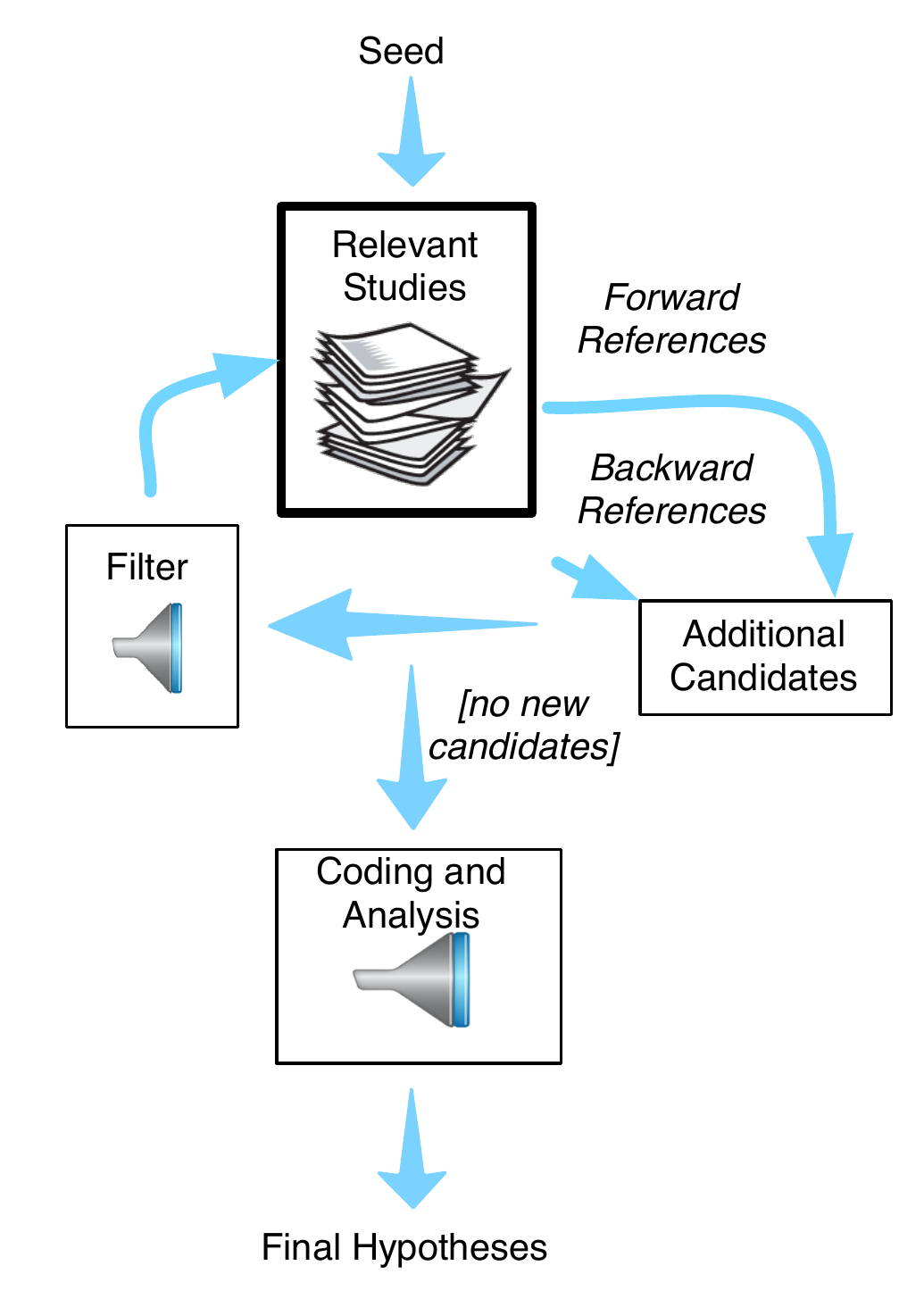}
 \caption{Overview of the literature search procedure.}
 \label{fig:search}
\end{wrapfigure}
We have repeated this process until we were unable to find additional relevant studies.
Afterwards, we analyzed this body of research, and used open coding to identify and group common patterns of results and conclusions.
This led to the formation of 15 hypotheses (categorized into 4 groups).
We illustrate our method in Figure~\ref{fig:search}.

\subsection{H1: Performance Predictability}

\newhypotheses

It is evident that cloud instances with different configurations (e.g., \texttt{m1.small} versus \texttt{m3.large} in EC2) provide different performance.  However, a plethora of existing research has established that even the performance of instances using the exact same configuration can vary dramatically~\cite{iosup:11,schad:10,cerotti:12}. This makes it hard for cloud customers to
predict the performance of their instances in advance. A wide array of existing research has shown that these variations are unique to virtualized cloud instances, or at least much less pronounced in on-premise environments~\cite{hazelhurst:08,kotthoff:14,hill:09,jackson:10,mehrotra:12,chakthranont:14}.

\hypothesis{inter}{\textbf{Performance Varies Between Instances} -- the performance of cloud instances using the same configuration tends to vary relevantly.}  

Existing literature supports \ref{hyp:inter} based on various micro and application benchmarks, including CPU performance~\cite{oLoughlin:13,akioka:10}, hard disk IO~\cite{kossmann:10,ghoshal:11,li:10}, memory~\cite{atas:14,gillam:13}, network latency~\cite{barker:10}, or overall system performance~\cite{walker:08}. However, the main causes of the observed performance variability differs for different types of benchmarks and applications. 
The performance of \emph{CPU-bound applications} or benchmarks (i.e., those that depend strongly on processing speed or memory access) varies primarily due to the randomness of which physical hardware is used for an instance (hardware heterogeneity). This has been studied (and exploited) extensively in~\cite{lenk:11,ou:13,farley:12,cerotti:12}.   

\hypothesis{hwd}{\textbf{CPU-Bound Applications} -- the performance of CPU-bound applications depends strongly on the served CPU model, and tends to vary primarily due to hardware heterogenity.} 

Contrary, the performance of \emph{IO-bound applications} or benchmarks (i.e., those that primarily depend on disk IO or networking bandwidth)
varies primarily due to the behavior of other co-located tenants~\cite{iosup:11}. That is, these applications
may suffer from a ``noisy neighbour'' effect~\cite{varadarajan:12}.

\hypothesis{mtd}{\textbf{IO-Bound Applications} -- the performance of IO-bound applications depends strongly on the behavior of other co-located tenants.}  

As a direct result, performance data collected from CPU-bound applications often follows a multi-modal distribution~\cite{schad:10,oLoughlin:13}. That is, while the overall performance variability is large, there are clusters of similarly-performing instances, which have been served with the same type of hardware. Hence, for CPU-bound applications, performance predictability
increases substantially if the hardware is known. For IO-bound applications, knowing the served hardware does not help much
to increase predictability. 


\subsection{H2: Variability Within Instances}

\newhypotheses

Additionally, even the performance of a single instance may vary over time. For IO-bound applications (e.g., disk IO~\cite{scheuner:14} and network throughput~\cite{wang:10,hill:10}), existing research has established that even multiple benchmarks taken in quick succession, within the timeframe of minutes or even seconds, can lead to substantially varying results.

\hypothesis{lifetime_mtd}{\textbf{Intra-Instance Variability of IO-Bound Applications} -- the performance of IO-bound applications tends to vary relevantly within the same instance.}

Theoretically, cloud providers are able to migrate instances from one physical host to another at runtime.
In practice, the hardware used to serve a given instance currently does not routinely
change after instance creation, leading to rather stable  performance of CPU-bound applications over time~\cite{dejun:09}. 

\hypothesis{lifetime_hw}{\textbf{Intra-Instance Stability of CPU-Bound Applications} -- the performance of CPU-bound applications tends not to vary relevantly within the same instance for instances with a dedicated CPU.}

An exception to \ref{hyp:lifetime_hw} are bursting cloud instance type families, such as GCE's \texttt{f1-micro} and EC2's \texttt{t1.micro}. 
Such
instance types do not receive an entire CPU, but share their processor with other tenants based on a credit system. Hence, their longer-term performance is also subject to noisy neighbours, as well as other performance variations induced by the credit
system\rev{~\cite{wen:15}}.

\hypothesis{lifetime_bursting}{\textbf{Intra-Instance Variability of Bursting Instance Types} -- the performance of any application using a bursting instance type tends to vary relevantly within the same instance.}

\subsection{H3: Temporal and Geographical Factors}

\newhypotheses

Some authors have speculated that there may be additional temporal and geographical factors that influence
cloud performance and predictability~\cite{iosup:11:2}, in addition to
hardware heterogeneity and multi-tenancy. One such potential factor of influence is the time of the
day when a benchmark was launched~\cite{borhani:14}.

\hypothesis{time}{\textbf{Impact of Time of the Day on Performance} -- the performance of a cloud instance in any application depends significantly on the time of the day.}

Moreover, we speculate that not only the absolute performance of cloud instances differs at different times of the
day, but also the performance variability.

\hypothesis{ptime}{\textbf{Impact of Time of the Day on Predictability} -- the predictability of the performance of a cloud instance in any application depends significantly on the time of the day.}

Other researchers have argued that the current day of the week may impact performance~\cite{lenk:11}.
The theory behind this, as well as behind \ref{hyp:time} and \ref{hyp:ptime}, is that enterprise applications are typically
used less during the weekend and outside of office hours, while other types of applications (e.g., movie
streaming services) are used more when the majority of their customers are not at work.

\hypothesis{day}{\textbf{Impact of Day of the Week on Performance} -- the performance of a cloud instance in any application depends significantly on the day of the week.}

Similarly to~\ref{hyp:ptime}, we assume that the predictability of the performance of a cloud instance also depends on the day.

\hypothesis{pday}{\textbf{Impact of Day of the Week on Predictability} -- the predictability of the performance of a cloud instance in any application depends significantly on the day of the week.}

Finally, multiple earlier studies have shown that the region an instance is launched in (e.g., \texttt{us-east-1} or \texttt{eu-west-1} in EC2)
has implications, both on the performance and the expected performance variability~\cite{dejun:09,schad:10}.
Different regions typically map to different physical data centers, which differ in the hardware used to serve requests, as well as in the data center utilization, which impacts how many tenants are typically located on each physical machine. To reflect this, regions are often not priced at equal rates.

\hypothesis{regions}{\textbf{Impact of Region on Performance} -- the performance of a cloud instance in any application depends significantly on the selected region.}

Again, we speculate that not only the absolute performance, but also the performance variability is impacted by the selected
provider region.

\hypothesis{pregions}{\textbf{Impact of Region on Predictability} -- the predictability of the performance of a cloud instance in any application depends significantly on the selected region.}

  

\subsection{H4: Instance Type Selection}

\newhypotheses

In H1 to H3, we have primarily focused on performance predictability, and what impacts it, in isolation.
However, practitioners choosing
between different configurations or providers are generally not interested in performance alone. A more
meaningful metric to select between instance types is the \emph{performance per US dollar per hour}~\cite{sobel:08}.
Clearly, the ratio of performance and costs is dependent on the used
application or benchmark. Hence, no consensus ``best'' instance type (or provider) for all use cases
has emerged from existing research. Nonetheless, we are able to identify a number of trends regarding the relationship
of different instance types.

One such trend is the observation that the ratio of performance and costs of more expensive instance types
tends to be lower than of cheaper instance types. Essentially, this means that there
are diseconomies of scale when selecting larger instances~\cite{borhani:14,exposito:13:2,sadooghi:15}. Note that this observation is
not obvious from provider specifications, which often give the impression that there is close to
a perfectly constant relationship between costs and expected performance, at least within the same instance type family.
Indeed, contemporary research sometimes seems to assume that twice as expensive instance types are generally twice as fast
(e.g.,~\cite{frincu:14}).

\hypothesis{diseconomies}{\textbf{Diseconomies of Scale of Larger Instance Types} -- the ratio of performance and costs for any application tends to decrease with increasing instance type costs.}

However, one advantage of larger instance types is a higher predictability of
performance~\cite{wang:11,kotthoff:14}.

\hypothesis{stability}{\textbf{Stability of Larger Instance Types} --  the predictability of performance for any application tends to increase with increasing instance type costs.} 

Finally, some existing studies have analyzed how specialized instance types (e.g., the IO-optimized instances from the ``i2'' family) compare to general-purpose instances. These
studies show that specialized instance types are more cost-efficient for the right kind of application~\cite{imai:13}.

\hypothesis{special}{\textbf{Price of Specialization} -- specialized instance types tend to have a better ratio of performance and cost for applications related to their specialization, and worse ratio otherwise.}

\section{Experimental Results}
\label{sec:experiments}

We now numerically discuss the formed hypotheses in the context of the performance of four existing IaaS providers.

\subsection{Data Collection}
\label{sec:data}

We used Cloud Workbench~\cite{scheuner:15} to set up and collect a relevant amount of performance measurements from
four real-life cloud providers. We have chosen EC2, GCE, Microsoft Azure, and IBM Softlayer (SL) for our study.
EC2 has been chosen due to both, its popularity in existing
benchmarking research and its high industrial relevance.
Similarly, Azure was chosen because it is seen as another current market leader~\cite{serrano:15}.
GCE, on the other hand, was interesting for us as the service went out of
beta only 1 month prior to the start of our data collection (on May 15th, 2014), and has, to the best of our knowledge, so
far only been used in a single performance study~\cite{li:13:3}.~\cite{serrano:15} lists GCE as a ``visionary'' in the current
cloud market. Finally, we have chosen SL as a smaller, less established, provider, who is nonetheless backed
by an established IT company.

Based on our literature review,
we decided to implement 3 micro-benchmarks, targeting \emph{instance processor speed} (CPU), (combined) \emph{disk read/write speed} (IO), and
(combined) \emph{memory read/write speed} (MEM). Additionally, we defined two application benchmarks, measuring \emph{queries per minute} on a MySQL database (OLTP)
and \emph{the Git checkout and Java compilation time of the open source project jCloudScale~\cite{leitner:12,zabolotnyi:15:1}} (Java). Following the distinction introduced in Section~\ref{sec:observations}, we consider the CPU, MEM, and Java benchmarks
to be CPU-bound, while IO and OLTP are IO-bound.   
More details on these benchmarks are given in Table~\ref{tab:benchmarks}. All EC2, Azure and SL benchmarks use an Ubuntu 14.04 LTS image as base operating system. In GCE, no
officially maintained Ubuntu images were available at the time of experimentation. Hence, we use Debian 7 Wheezy base images for this provider. 

\begin{wraptable}{r}{0.65\textwidth}
\centering
\scriptsize
\begin{tabular}{|c|c|p{5cm}|c|}
	\hline
   \rowcolor[HTML]{EFEFEF} & \textbf{Name} & \textbf{Benchmark Description} & \textbf{Unit}\\
  \hline
  \multirow{11}{*}{\rotatebox[origin=c]{90}{CPU-Bound}} & CPU & Uses \texttt{sysbench} to check 20000 natural numbers for prime-ness; measures time to completion; lower is better & s \\
  \cline{2-4}
  &  MEM & Uses \texttt{mbw} to 50 times allocate two 64 MiB arrays in memory and copy one array to the other; measures average memory bandwidth; higher is better &  MiB/s \\
\cline{2-4}
  &  Java & Clones jCloudScale, a mid-sized Java project from GitHub and compiles it using OpenJDK 7; measures end-to-end duration; lower is better & s \\   
  \hline
  \multirow{7}{*}{\rotatebox[origin=c]{90}{IO-Bound}}  & IO & Uses \texttt{sysbench} to repeatedly read / write a 5 GByte file for 3 minutes; measures average combined disk read/write bandwidth; higher is better & Mb/s \\
  \cline{2-4}
   & OLTP & Uses \texttt{sysbench} to repeatedly execute SQL queries against a MySQL database with 10.000 rows for 3 minutes; measures average queries per second; higher is better & Query/s \\   
  \hline
\end{tabular}
\caption{Benchmarks used for data collection.}
\label{tab:benchmarks}
\end{wraptable}
Further, we considered two types of experiments.
In \emph{isolated tests}, we
acquire a specific instance type from a provider, provision the
benchmark code (e.g., install the \texttt{sysbench} tool for the CPU
benchmark), execute the benchmark three times directly after each other, and
then immediately release the instance. Using Cloud Workbench, we
scheduled this procedure in 82 configurations 6 times per day over a period of
approximately one month during July/August 2014 for EC2 and GCE, and 
during June/July 2015 for Azure and SL.
For European regions (``eu''), we used \texttt{eu-west-1} in case of EC2, \texttt{West Europe}
for Azure, and \texttt{europe-west1-a} in case of GCE. For North American regions (``na''),
we used \texttt{us-east-1} for EC2, \texttt{East US} for Azure, \texttt{mel01} (Montreal)
in case of SL, and
\texttt{us-central1-a} for GCE.

In some cases, we had to deal with transient faults, which originated either in the
cloud provider (e.g., instances not starting up correctly) or in our own tooling
(e.g., timeouts from the Chef server used for instance provisioning).
In these
cases, we have simply cancelled the benchmarking run and discarded the resulting
measurements.
As executing all combinations of benchmarks, providers, instance types, and regions
proved prohibitively expensive, we aimed for a pragmatic compromise between cost- and time-efficient
benchmarking on the one hand, and validity and expressiveness of the resulting data on the other.
Specifically, we executed the OLTP
and MEM benchmarks only in EC2.
\rev{For OLTP, the main reason was that we saw a strong correlation between OLTP and IO. Hence,
executing both benchmarks in all clouds was deemed not necessary. The same was true for MEM and CPU,
as both benchmarks tend to depend strongly on the served CPU model.}
Similarly, we have only analyzed IO- and CPU-optimized instance types
in EC2 and in the European region.
SL does not specifically provide named instance types. Rather, customers can specify various virtualized hardware properties, such as number of virtual CPUs and available memory. As SL does not provide an equivalent to the bursting micro instance types of other providers, we only benchmark two configurations, resembling small and large instance types in other clouds. Further, we have only benchmarked SL in North America.

Orthogonal to these isolated tests, we also collected data related to the performance stability of cloud virtual machines
over time.
For these \emph{continuous tests}, we again acquire a specific type of virtual resource
from a provider and provision the benchmark.
However, we then continuously execute the benchmark once every hour for a
total duration of 3 days (72 executions), after which we discard the instance. For each benchmarked
configuration, we repeated this 15 times. We executed all micro-benchmarks used in our study (CPU, IO and MEM)
for these experiments.
Further, we used European regions for EC2, GCE, and Azure, and the Montreal region for Softlayer.
Table~\ref{tab:data} lists the number of collected data points for each cloud provider and data set.



\subsection{Validation Results}

We now discuss the validation of the 15 hypotheses formed in Section~\ref{sec:observations}.
We base our validation on the data described in Section~\ref{sec:data}.

\subsubsection{Performance Predictability of Instances}

We first validate the hypotheses related to the performance variability between instances acquired with identical configuration.
In this context, a configuration $c \in \mathcal{C}$ is defined as the quadruple of cloud provider, region, instance type, and benchmark $\langle \mathcal{P}, \mathcal{R}, \mathcal{T}, \mathcal{B} \rangle$.
\begin{wraptable}{r}{0.55\textwidth}
\centering
\scriptsize
\begin{tabular}{|l|c|c|c|c|c|}
\hline
\rowcolor[HTML]{EFEFEF} & EC2 & GCE & Azure & SL & \textbf{Total} \\
\hline
 Isolated &  21552 & 8372 & 9406 & 3041 & 42371 \\
\hline
 Continuous &  3243 & 3225 & 2995 & 2084 & 11547 \\
 \hline
\textbf{Total} &  24795 & 11597 & 12401 & 5125 & 53918 \\
 \hline 
\end{tabular}
\caption{\rev{Overall} number of collected measurements per cloud provider.}
\label{tab:data}
\end{wraptable}
Every configuration has associated instances $i(c) \in \mathcal{I}$ (where $\mathcal{I}$ is the set of all instances). For every instance within a configuration, we collected performance measurements $m(i) \in \{m_1, \dots, m_n \}, \textit{ where } m_1, \dots, m_n \in \mathbb{R}$. 
For each configuration, we collected a large number of individual measurements (approximately 500 measurements each, resulting in 42371 data points), consisting of the union of all
measurements of instances of this configuration (Equation \ref{eq:m_c}).
 
\begin{equation}
\label{eq:m_c}
	m_c = \bigcup_{i \in i(c)} m(i)
\end{equation}

As a measure of performance variability, we use the \emph{relative standard
deviation} ($c_{RSD}$) of the measurements collected for each configuration, as defined in Equation~\ref{eq:rsd}, with $\overline{m_c}$ referring to the arithmetic mean of $m_c$, and $\sigma_{m_c}$ referring to the standard deviation of $m_c$.


\begin{equation}
\label{eq:rsd}
\forall c \in \mathcal{C} \; : \; c_{RSD} = 100 \frac{\sigma_{m_c}}{\overline{m_c}}
\end{equation}

We display $c_{RSD}$ of all configurations in Table~\ref{tab:inter}. We assume that a relative standard deviation of more than 5\% represents a \emph{relevant} variability in performance for most use cases. However, evidently, specific applications may tolerate higher or lower variability.

\begin{table}[h!]
\centering
\scriptsize
\begin{tabular}{|l|l|l|lll|ll|}
\hline
\rowcolor[HTML]{EFEFEF} & & & \multicolumn{3}{c|}{\textbf{CPU-Bound}} & \multicolumn{2}{c|}{\textbf{IO-Bound}}\\
\hline
\rowcolor[HTML]{EFEFEF} &  & \textbf{Type} & \textbf{CPU}  & \textbf{MEM} & \textbf{Java}  & \textbf{IO} & \textbf{OLTP} \\
\hline
\multirow{8}{*}{\rotatebox[origin=c]{90}{EC2}} & \multirow{5}{*}{\rotatebox[origin=c]{90}{eu}}  & t1.micro & \cellcolor{red!6} 12.14 &  \cellcolor{red!9}17.67 & \cellcolor{red!15} 30.63 & \cellcolor{red!35} 71.33 & \cellcolor{red!15} 30.66 \\
\cline{3-3}
 &  & m1.small & \cellcolor{red!2} 3.19 & \cellcolor{red!2} 3.77& \cellcolor{red!1} 3.17 & \cellcolor{red!44} 88.49 & \cellcolor{red!6} 13.02 \\
\cline{3-3}
&  & m3.large &\cellcolor{white!1} 0.13 & \cellcolor{red!1} 2.07 & \cellcolor{red!4} 7.22 & \cellcolor{red!18} 35.53 & \cellcolor{red!11} 21.26\\
\cline{3-3} 
&  & c3.large & \cellcolor{white!1} 0.21 & \cellcolor{red!4} 8.60& \cellcolor{red!3} 6.42 & \cellcolor{red!29} 58.88 & \cellcolor{red!11} 21.31\\
\cline{3-3} 
&  & i2.xlarge & \cellcolor{white!1} 0.12 & \cellcolor{red!6} 11.92 & \cellcolor{red!4} 8.44 & \cellcolor{red!10} 20.07 & \cellcolor{red!6} 12.28 \\
\cline{2-3} 
& \multirow{3}{*}{\rotatebox[origin=c]{90}{na}}  & t1.micro & \cellcolor{red!10} 20.28 & \cellcolor{red!13} 26.40 & \cellcolor{red!30} 59.32 & \cellcolor{red!35} 70.08 & \cellcolor{red!16} 32.18 \\
\cline{3-3}
 &  & m1.small & \cellcolor{red!6} 12.81 & \cellcolor{red!13} 26.18 & \cellcolor{red!3} 5.34 & \cellcolor{red!47} 94.47 & \cellcolor{red!8} 15.68 \\
 \cline{3-3}
&  & m3.large & \cellcolor{white!1} 0.16 & \cellcolor{red!2} 4.46 & \cellcolor{red!5} 9.23 & \cellcolor{red!25} 49.02 & \cellcolor{red!18} 37.10 \\
\hline 
\multirow{6}{*}{\rotatebox[origin=c]{90}{GCE}} & \multirow{3}{*}{\rotatebox[origin=c]{90}{eu}} & f1-micro & \cellcolor{red!3} 5.28 & \cellcolor{gray!25}  & \cellcolor{red!4} 8.36 &  \cellcolor{red!2} 3.06 & \cellcolor{gray!25} \\
\cline{3-3}
 &  & n1-standard-1 & \cellcolor{red!1} 2.54 & \cellcolor{gray!25}  & \cellcolor{red!3} 6.99 &  \cellcolor{red!2} 3.36 & \cellcolor{gray!25} \\
 \cline{3-3}
 &  & n1-standard-2 & \cellcolor{red!1} 1.71 & \cellcolor{gray!25}  & \cellcolor{red!3} 6.96 &  \cellcolor{red!1} 1.33 & \cellcolor{gray!25} \\
 \cline{2-3}
 & \multirow{3}{*}{\rotatebox[origin=c]{90}{na}} & f1-micro & \cellcolor{red!3} 5.13 & \cellcolor{gray!25}  & \cellcolor{red!4} 7.17 &  \cellcolor{red!5} 9.47 & \cellcolor{gray!25} \\
 \cline{3-3}
 &  & n1-standard-1 & \cellcolor{red!1} 2.05  & \cellcolor{gray!25}  & \cellcolor{red!4} 8.31 &  \cellcolor{red!5} 10.39 & \cellcolor{gray!25} \\
 \cline{3-3}
 &  & n1-standard-2 & \cellcolor{red!1} 1.16  & \cellcolor{gray!25}  & \cellcolor{red!5} 9.53 &  \cellcolor{red!2} 4.88 & \cellcolor{gray!25} \\
 \hline
\multirow{6}{*}{\rotatebox[origin=c]{90}{Azure}} & \multirow{3}{*}{\rotatebox[origin=c]{90}{eu}} & ExtraSmall & \cellcolor{red!9} 18.38 & \cellcolor{gray!25}  & \cellcolor{red!8} 16.88 &  \cellcolor{red!30} 61.92 & \cellcolor{gray!25} \\
\cline{3-3}
 &  & Small & \cellcolor{red!9} 18.23 & \cellcolor{gray!25}  & \cellcolor{red!4} 8.37 &  \cellcolor{red!30} 59.01 & \cellcolor{gray!25} \\
 \cline{3-3}
 &  & Medium & \cellcolor{red!9} 17.81 & \cellcolor{gray!25}  & \cellcolor{red!6} 11.91 &  \cellcolor{red!24} 47.14 & \cellcolor{gray!25} \\
 \cline{2-3}
 & \multirow{3}{*}{\rotatebox[origin=c]{90}{na}} & ExtraSmall & \cellcolor{red!9} 18.13 & \cellcolor{gray!25}  & \cellcolor{red!8} 15.96 &  \cellcolor{red!25} 49.01 & \cellcolor{gray!25} \\
 \cline{3-3}
 &  & Small & \cellcolor{red!10} 19.11  & \cellcolor{gray!25}  & \cellcolor{red!3} 6.62  & \cellcolor{red!22} 44.01 & \cellcolor{gray!25} \\
 \cline{3-3}
 &  & Medium & \cellcolor{red!9} 18.28  & \cellcolor{gray!25}  & \cellcolor{red!5} 10.96 &  \cellcolor{red!24} 48.31 & \cellcolor{gray!25} \\
 \hline 
\multirow{2}{*}{\rotatebox[origin=c]{90}{SL}} & \multirow{2}{*}{\rotatebox[origin=c]{90}{na}} & 1 CPU / 2048 MB & \cellcolor{white!1} 0.11 & \cellcolor{gray!25}  & \cellcolor{red!3} 6.65 &  \cellcolor{red!7} 13.01 & \cellcolor{gray!25} \\
\cline{3-3}
 &  & 2 CPUs / 4096 MB & \cellcolor{white!1} 0.11 & \cellcolor{gray!25}  & \cellcolor{red!4} 7.14 & \cellcolor{red!3} 6.27 & \cellcolor{gray!25} \\
 \hline  
\end{tabular}
\caption{$c_{RSD}$ of all configurations in isolated tests. All values in \%.}
\label{tab:inter}
\end{table}             

We note four observations. Firstly, in 63 of 82 configurations (i.e., unique combinations
of provider, region, instance type, and benchmark), we
experienced a $c_{RSD}$ of more than 5\%. Hence, we argue that the fundamental
hypothesis \ref{hyp:inter} is supported by our data.
Secondly, there is a clear difference between  the analyzed providers.
IO (and, to a lesser
extent, OLTP)
performance in EC2, as well as IO in Azure, seems almost chaotic, while the same benchmarks are much more
predictable in GCE and SL.
CPU performance is generally rather predictable with the exception of Azure, for which
all benchmarks including CPU are highly unpredictable.
By and large, performance in GCE and SL is much more predictable than in EC2 and Azure
at the time of our experiments. Thirdly, EC2 \texttt{t1.micro} instances in the \texttt{us-east-1}
region have been the most unpredictable type of cloud virtual machine in our experiments,
with a $c_{RSD}$ larger than 20\% in all benchmarks.
Fourthly, the two benchmarks we classify as IO-bound (IO and OLTP) are substantially less predictable than those where performance is strongly influenced by the served hardware (CPU, MEM, and Java) in EC2 and Azure. In SL and GCE,
this difference is much less pronounced. 


These results can be explained by analyzing the CPU models we have received from the cloud.
Contrary to our expectations, we have not universally observed hardware heterogeneity in our experiments.
Virtual machines in GCE currently report a standard
Intel CPU independent of the actual hardware (and our measured performance data makes it
plausible that the underlying physical CPUs are indeed very similar or identical).
Virtual machines in EC2 report the actual
hardware that the instance is running on. However, as assured by AWS,
\texttt{m3.large} instances always receive an Intel
Xeon E5-2670 CPU, \texttt{c3.large} instance always receive an Intel Xeon E5-2680 CPU, and \texttt{i2.xlarge}
instance always
receive an Intel Xeon E5-2670 v2 CPU. Hence, only the \texttt{t1.micro} and \texttt{m1.small} instances are theoretically
prone to
experience
hardware heterogeneity in this provider, and even for  \texttt{m1.small} instances in the \texttt{eu-west-1} region
we have in practice seen the exact same CPU model in more than 97\% of all runs.
SL also provides well-defined hardware for each instance type.
Only Azure actually uses a wider range of CPU models for all analyzed instance types.  
Given that \ref{hyp:hwd} postulates that, for CPU-bound applications,
hardware heterogeneity accounts for most of the observed performance variability,
the apparent \emph{lack} of heterogeneity in practice explains why the CPU-bound
benchmarks ended up with relatively little variability everywhere except in the Azure cloud.  


To assess \ref{hyp:hwd} and \ref{hyp:mtd}, we now look in detail at a subset of those instance types for which
hardware heterogenity was indeed a factor, concretely EC2 \texttt{m1.small} and the Azure \texttt{Small} configurations
in the North American regions.
In Table~\ref{tab:hw}, we provide another table of relative standard deviations for those configurations. However,
this time, we additionally control for the concrete CPU model we have received in our experiments. To calculate the proper relative standard deviation (Equation \ref{eq:cpu_rsd}), we consider measurements for a certain hardware type ($cpu \in CPU(c)$, Table \ref{tab:types}) within a configuration $c \in \mathcal{C}$: $m_{c,cpu} \in m_c$. The column ``\#'' indicates
the number of concrete benchmark runs in this configuration that have received this CPU model from the cloud provider.

\begin{equation}
\label{eq:cpu_rsd}
\forall c \in \mathcal{C} \; \forall cpu \in CPU(c) \; : \; cpu_{c,RSD} = 100 \frac{\sigma_{m_{c, cpu}}}{\overline{m_{c, cpu}}}
\end{equation}

Our data indeed shows that controlling for the CPU model further reduces the
observed relative standard
deviations in CPU-bound benchmarks, supporting~\ref{hyp:hwd}. Controlling
for the served CPU model also reduces the relative standard
deviations in IO-bound benchmarks, but to a much smaller degree. Hence,
we consider \ref{hyp:mtd} to also be supported by our results.

\begin{table}[h!]
\centering
\scriptsize
\begin{tabular}{|l|l|c|lll|ll|}
\hline
\rowcolor[HTML]{EFEFEF} & & & \multicolumn{3}{c|}{\textbf{CPU-Bound}} & \multicolumn{2}{c|}{\textbf{IO-Bound}}\\
\hline
\rowcolor[HTML]{EFEFEF} & \textbf{Model} & \textbf{\#} & \textbf{CPU}  & \textbf{MEM} & \textbf{Java}  & \textbf{IO} & \textbf{OLTP} \\
\hline
		\multirow{6}{*}{\rotatebox[origin=c]{90}{EC2}} & Intel E5-2650 & 1962 & \cellcolor{red!1} 0.42 & \cellcolor{red!1} 1.53 & \cellcolor{red!3}4.76 & \cellcolor{red!44} 88.4 & \cellcolor{red!7} 14.87 \\
		\cline{2-3}
		 & Intel E5430 & 364 & \cellcolor{red!1} 0.15 & \cellcolor{red!3} 4.51 & \cellcolor{red!3} 4.84 & \cellcolor{red!16}31.53 & \cellcolor{red!3} 6.56 \\
		 \cline{2-3}
		 & Intel E5645 & 293 & \cellcolor{red!1} 0.36 & \cellcolor{red!2} 4.6 & \cellcolor{red!3} 5.04 & \cellcolor{red!11} 22.82 & \cellcolor{red!5} 8.86 \\
		 \cline{2-3}
		 & Intel E5507 & 84 & \cellcolor{red!1} 0.25 & \cellcolor{red!2} 2.66 & \cellcolor{red!1} 2.79 & \cellcolor{red!11} 21.02 & \cellcolor{red!4} 8.51 \\
		 \cline{2-3}
		 & Intel E5-2651 & 32 & \cellcolor{red!1} 0.28 & \cellcolor{red!1} 0.54 & \cellcolor{red!2} 3.03 & \cellcolor{red!1} 2.27 & \cellcolor{red!9} 18.27 \\
		 \cline{2-3}
		 & AMD 2218 HE & 9 & \cellcolor{gray!25} & \cellcolor{gray!25} & \cellcolor{gray!25} & \cellcolor{red!3} 5.45 & \cellcolor{red!1} 0.98 \\
		\hline

		 \multirow{3}{*}{\rotatebox[origin=c]{90}{Azure}} & AMD 4171 HE & 782 & \cellcolor{red!1} 1.84 & \cellcolor{gray!25} & \cellcolor{red!3} 5.8 & \cellcolor{red!25} 49.05 & \cellcolor{gray!25}  \\
		\cline{2-3}
		 & Intel E5-2673 & 595 & \cellcolor{red!2} 3.05 & \cellcolor{gray!25} & \cellcolor{red!2} 4.13 & \cellcolor{red!15} 29.85 & \cellcolor{gray!25}  \\
		\cline{2-3}
		 & Intel E5-2660 & 189 & \cellcolor{red!1} 0.8 & \cellcolor{gray!25} & \cellcolor{red!2} 3.46 & \cellcolor{red!24} 47.51 & \cellcolor{gray!25}  \\		
		\hline

	\end{tabular}
		\caption{$cpu_{c,RSD}$ for \texttt{m1.small} EC2 instances and Azure \texttt{Small} instances in isolated tests in the North American regions. All values are in \%.}
		\label{tab:hw}
\end{table}

The other configurations where we observed hardware heterogenity (all other Azure
configurations, the \texttt{t1.micro} configurations in EC2, and, to a very small degree,
the \texttt{m2.small} configuration in Europe) lead to comparable results. The detailed
data for these cannot be shown here for reasons of brevity, but this data is available
in our online appendix.

%
%

\subsubsection{Performance Variability Within Instances}

So far, we have evaluated the variability of performance data measured on
different cloud instances.
Now we turn to the  variability of performance within a single instance. We
consider 33 different continuous configurations $c_l \in \mathcal{C}_l$.
For each continuous configuration, we requested 15 instances $i(c_l) \in \mathcal{I}$, and
executed the benchmark once per hour for a total duration of 72 hours, leading to a
series of measurements per instance $m(i) \in \{m_1, \dots, m_n \}, \textit{ where } m_1, \dots, m_n \in \mathbb{R}$.
We again use the relative standard deviation ($c_{l_{RSD}}$, analogously defined as in Equation \ref{eq:rsd}) as a measure of performance
variability, this time of the variability within a single cloud instance
(Equation~\ref{eq:rsd2}).  In addition, we define the \emph{mean relative standard deviation}
($\overline{c_{l_{RSD}}}$) as the arithmetic mean of all $i_{RSD}$ values for a
continuous configuration over all 15 instances that were provisioned using this configuration.

\begin{equation}
\label{eq:rsd2}
\forall c_l \in \mathcal{C}_l \: \forall i \in c_l \; : \; i_{RSD} = 100 \frac{\sigma_{m(i)}}{\overline{m(i)}}
\end{equation}

\begin{figure}[h]
\centering
\begin{minipage}{0.45\textwidth}
\includegraphics[width=\linewidth]{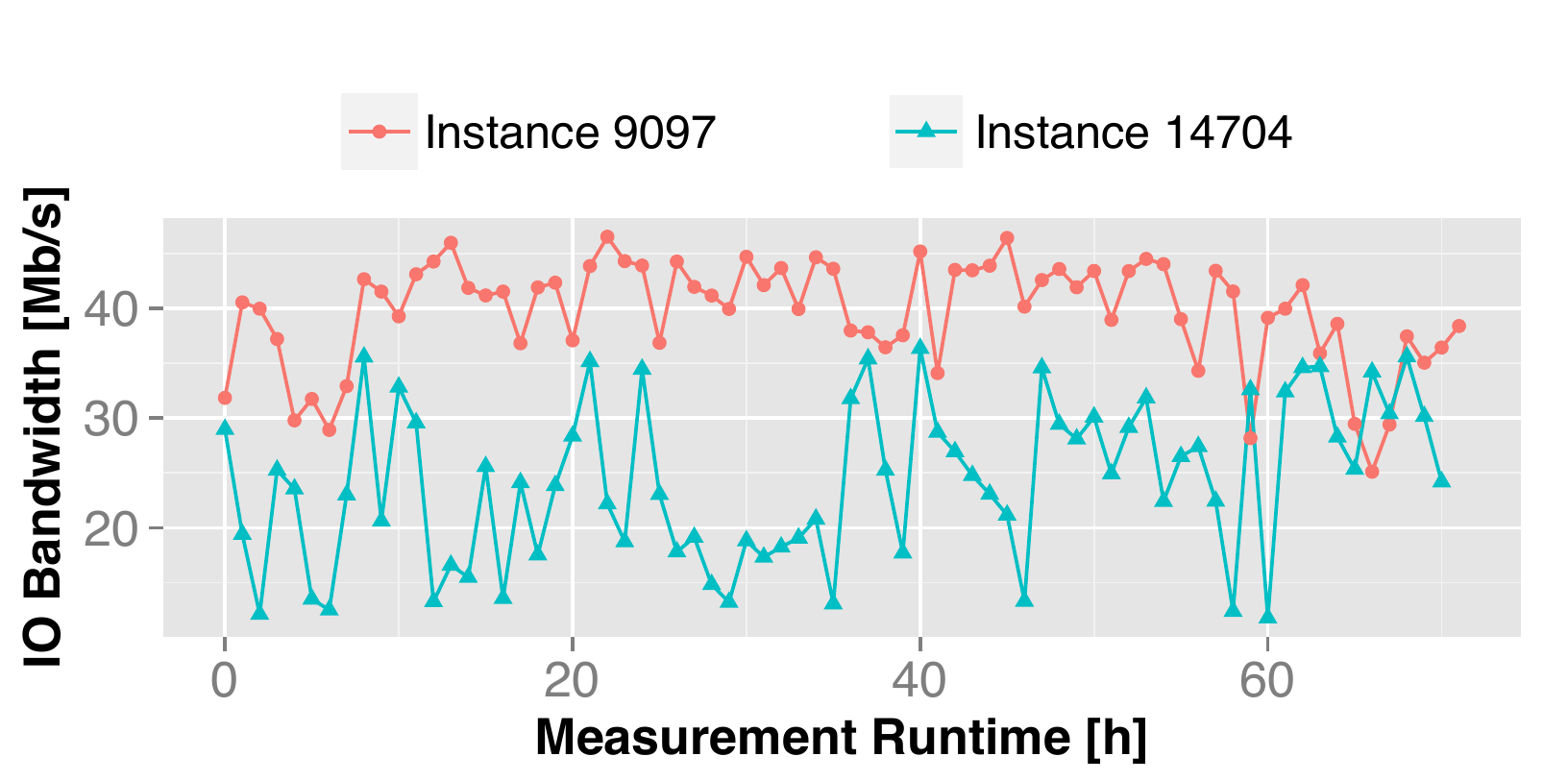}
 \caption{IO measurements on 2 example instances in EC2's \texttt{eu-west-1} region. Both instances are \texttt{m3.large} instance types.}
 \label{fig:io_per_time}
\end{minipage}
\hfill
\begin{minipage}{0.45\textwidth}
 \includegraphics[width=\linewidth]{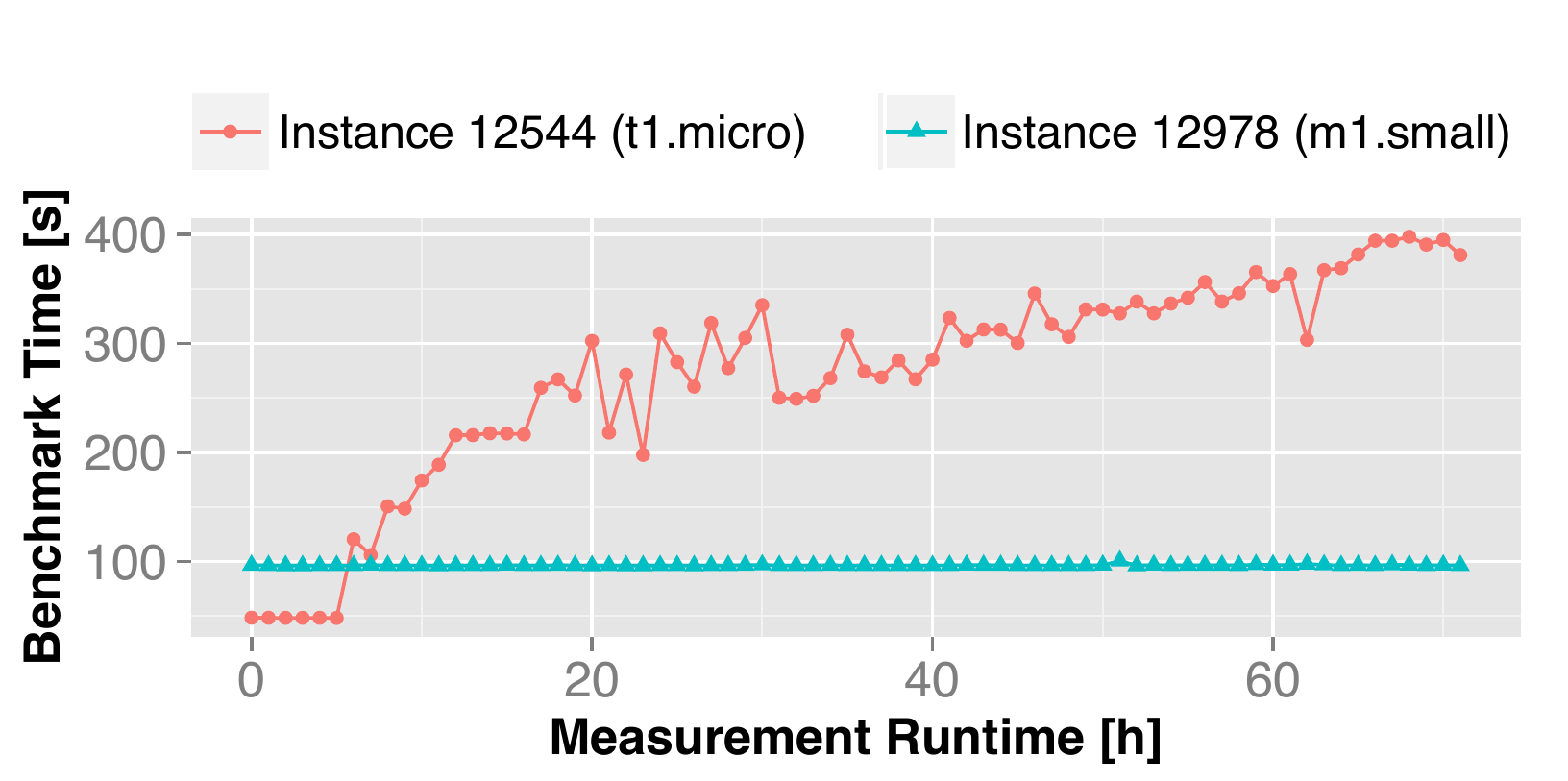}
 \caption{CPU measurements on 2 instances in EC2's \texttt{eu-west-1} region. 12544 is \texttt{t1.micro}, and 12978 is a \texttt{m1.small} instance.}
 \label{fig:cpu_per_time}
 \end{minipage}
\caption{Performance Variability Within Instances}
\label{fig:basic}
\end{figure}

Following \ref{hyp:lifetime_mtd}, we expect IO-bound benchmarks
to exhibit relevant performance variability even for measurements on the same
instance. We initially explored this visually by plotting measurements of different
instances on a time line. This visual data analysis supported our hypothesis.
Further, we observe that there are substantial differences between individual instances.
That is, there are  slow and fast, as well as stable and unstable instances.
This is exemplified for two sample instances using the same
configuration (EC2 in the \texttt{eu-west-1} region, using the \texttt{m3.large} instance type,
and testing IO performance) in Figure~\ref{fig:io_per_time}. In  this example,
despite identical configuration, the mean IO read/write speed of instance 9097 is almost
twice as high as of instance 14704. Further, the performance of 14704 fluctuates
more ($i_{RSD} = 12.25\%$ for 9097 versus $i_{RSD} = 30.12\%$ for 14704). However,
there is no obvious trend in either of the time series (i.e., the instances do not
seem to slow down or get faster over time).

Contrary, \ref{hyp:lifetime_hw} states that CPU-bound benchmarks do not
exhibit relevant performance variability for measurements on the same
instance, as long as the instance is \emph{not} using a bursting instance
type (the latter exception is captured by \ref{hyp:lifetime_bursting}).

We again verified these hypotheses using  visual exploratory data analysis.
Figure~\ref{fig:cpu_per_time} plots the CPU measurements of two examples instances, this time using different
instance types (12544, \texttt{t1.micro}, and 12978, \texttt{m1.small}). Note that, for the CPU benchmark, lower values 
represent higher performance.

We observe that the \texttt{m1.small} instance 12978 has essentially no variability of CPU
performance over time  
($i_{RSD} = 0.57\%$). The bursting instance 12544 (with a shared CPU) not only has a substantial variability ($i_{RSD} = 34.82\%$),
but performance is actually decreasing over time. This is because we did not give
the bursting instance time to replenish its credit account during our tests. Hence, the CPU cycles available to the
instance gradually depleted over time, leading to the observed slowdown. However, it is interesting to see that
the cheaper \texttt{t1.micro} instance actually starts off \emph{faster} than the more expensive \texttt{m1.small}
until its credit account starts to deplete.

%


To analyze \ref{hyp:lifetime_mtd}, \ref{hyp:lifetime_hw}, and \ref{hyp:lifetime_bursting} more formally, we
now provide the mean relative standard deviations $\overline{c_{l_{RSD}}}$ for each continuous configuration in Table~\ref{tab:intra}.
\begin{wraptable}{r}{0.55\textwidth}
\centering
\scriptsize
\begin{tabular}{|l|l|l|ccc|}
\hline
\rowcolor[HTML]{EFEFEF} &  & \textbf{Type} & \textbf{CPU}  & \textbf{MEM} & \textbf{IO} \\
\hline
\multirow{3}{*}{\rotatebox[origin=c]{90}{EC2}} & \multirow{3}{*}{\rotatebox[origin=c]{90}{eu}}  & t1.micro & \cellcolor{red!40} 40.17 & \cellcolor{red!40} 40.86 & \cellcolor{red!22} 22.78 \\
\cline{3-3}
 &  & m1.small & \cellcolor{red!2} 1.69 & \cellcolor{red!2} 1.66 & \cellcolor{red!27} 27.28 \\
 \cline{3-3}
&  & m3.large & \cellcolor{red!1} 0.89 & \cellcolor{red!1} 0.49 & \cellcolor{red!18} 17.92 \\
\hline 
\multirow{3}{*}{\rotatebox[origin=c]{90}{GCE}} & \multirow{3}{*}{\rotatebox[origin=c]{90}{eu}} & f1-micro &  \cellcolor{red!3} 3.24 & \cellcolor{red!7} 6.85 &  \cellcolor{red!3} 2.65 \\
\cline{3-3}
 &  & n1-standard-1 & \cellcolor{red!1} 0.76 & \cellcolor{red!2} 2.43  &  \cellcolor{red!4} 4.42 \\
 \cline{3-3}
 &  & n1-standard-2 & \cellcolor{red!1} 0.86 & \cellcolor{red!2} 2.01  &  \cellcolor{red!1} 1.46 \\
 \hline
\multirow{3}{*}{\rotatebox[origin=c]{90}{Azure}} & \multirow{3}{*}{\rotatebox[origin=c]{90}{eu}} & ExtraSmall &  \cellcolor{red!1} 2.41 & \cellcolor{red!2} 3.39 &  \cellcolor{red!14} 27.08 \\
\cline{3-3}
 &  & Small & \cellcolor{red!1} 1.65 & \cellcolor{red!1} 2.33  &  \cellcolor{red!47} 93.47 \\
 \cline{3-3}
 &  & Medium & \cellcolor{red!1} 1.30 & \cellcolor{red!1} 1.98  &  \cellcolor{white!1} 0.14 \\
 \hline 
\multirow{2}{*}{\rotatebox[origin=c]{90}{SL}} & \multirow{2}{*}{\rotatebox[origin=c]{90}{na}} & 1 CPU / 2048 MB &  \cellcolor{white!1} 0.22 & \cellcolor{white!1} 0.39 &  \cellcolor{red!1} 2.94 \\
\cline{3-3}
 &  & 2 CPU / 4096 MB & \cellcolor{white!1} 0.13 & \cellcolor{white!1} 0.14  &  \cellcolor{red!1} 2.25 \\
 \hline 
  
\end{tabular}
\caption{$\overline{c_{l_{RSD}}}$ of all configurations in continuous tests. All values are in \%.}
\label{tab:intra}
\end{wraptable}   
We observe that, as expected, the relative standard deviation within the same cloud instances is very low
for all CPU-bound benchmarks on non-bursting instance types in all providers. In terms of IO, Azure and EC2
exhibit fluctuating performance for most instance types, while both, GCE and SL, are remarkably predictable.
Hence, we conclude that \ref{hyp:lifetime_hw} is supported for all providers, while \ref{hyp:lifetime_mtd}
is only supported in EC2 and Azure. Further, note that, with the exception of the \texttt{Small} instance type in
Azure, all configurations have less intra-instance performance variability than between different instances
of the same configuration.
The performance of the \texttt{t1.micro} instance type on EC2
varies wildly, even (and especially) for CPU-bound benchmarks.
Performance in GCE is again remarkably predictable, even for bursting instance types.
Hence, we conclude that our data supports \ref{hyp:lifetime_bursting} only for EC2.

\subsubsection{Temporal and Geographical Factors}

\begin{figure}[h]
\centering
 \includegraphics[width=0.75\linewidth]{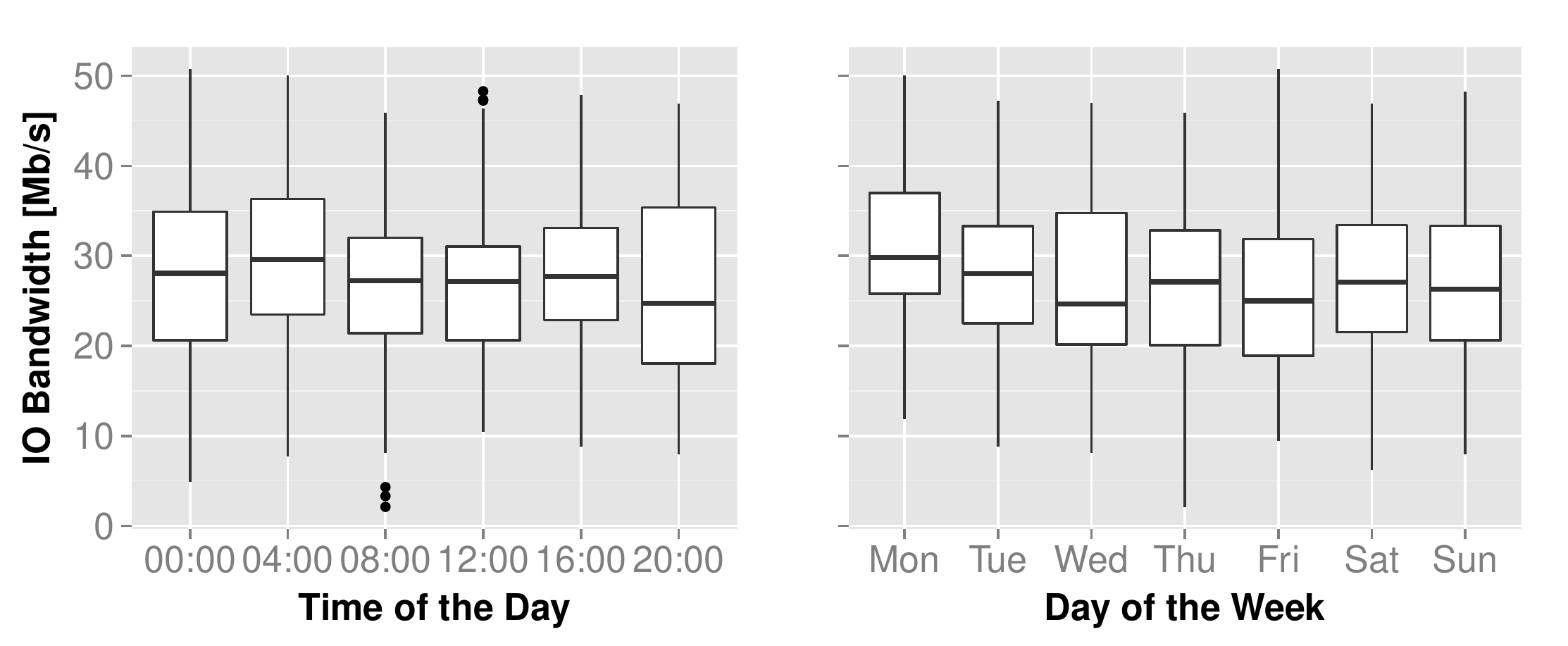}
 \caption{Distributions of IO performance per time of the day and day of the week for \texttt{m3.large} EC2 instances in the \texttt{eu-west-1} region.}
 \label{fig:datetime}
\end{figure}

In \ref{hyp:time}-\ref{hyp:pregions}, we speculate that time of day, day of the week, and the region are influential factors of overall performance and predictability. To this end, we have grouped the data collected from the isolated tests discussed in Section~\ref{sec:data} into days (Monday through Sunday) and time slots (in 4-hour slots, i.e., slot 1 is from ``00:00 AM'' to ``03:59 AM''), based on the start time of the benchmark. The resulting measurements
are approximately equally distributed over days and time slots. The
region information is readily available, as it is part of the configuration $c \in \mathcal{C}$.   
Using visual data analysis, we were, against our expectations, not able to observe any strong indication that would support an impact of the time of the day or the day of the week, either on absolute performance or predictability. For illustration, Figure~\ref{fig:datetime} depicts the distributions of IO performance for \texttt{m3.large} instances in \texttt{eu-west-1} in standard Boxplot notation. The remaining configurations show similar distributions.

%
%

To analyze this more formally, we now test for statistically significant differences in the underlying
distributions. We are required to perform pairwise comparisons
for each pair of time slots or days, and each configuration. That means that for each $c \in \mathcal{C}$ we need to evaluate 21 p-values
respectively for \ref{hyp:day} and \ref{hyp:pday} (measurements on Mondays versus measurements on Tuesdays, Mondays versus Wednesdays, etc.), and 15 p-values for each of \ref{hyp:time} and \ref{hyp:ptime}.
Hence, we ended up with a total of 3444 p-values for days of the week, and 2460 p-values for time slots. 
Given that our data is not normally distributed, we performed Wilcoxon signed-rank tests rather than standard Student's t-tests.
P-values were adjusted for multiple hypotheses testing using Holm correction. We tested for statistically significant differences in performance variability via Levene's variance test. The assumed null-hypothesis for all tests was that there is \emph{no} difference in underlying distributions between different days or time slots. We tested for a confidence level of $0.01$.


\begin{table}[h!]
	\scriptsize
      \centering
	\begin{tabular}{|l|cc|cc|l|cc|cc|}
		\hline
		\rowcolor[HTML]{EFEFEF} \multicolumn{5}{|l|}{\textbf{Time of the Day}} & \multicolumn{5}{|l|}{\textbf{Day of the Week}} \\
		\hline
		\rowcolor[HTML]{EFEFEF}  & \multicolumn{2}{c|}{\textbf{\ref{hyp:time}}} & \multicolumn{2}{c|}{\textbf{\ref{hyp:ptime}}} & & \multicolumn{2}{c|}{\textbf{\ref{hyp:day}}} & \multicolumn{2}{c|}{\textbf{\ref{hyp:pday}}} \\
		\rowcolor[HTML]{EFEFEF}  & $p < 0.01$ & $p \ge 0.01$ & $p < 0.01$ & $p \ge 0.01$ & & $p < 0.01$ & $p \ge 0.01$ & $p < 0.01$ & $p \ge 0.01$ \\				
		\hline
		00:00 & \cellcolor{ForestGreen!2}4\% & \cellcolor{red!48}96\%  & \cellcolor{ForestGreen!12}24.4\%  & \cellcolor{red!38}75.6\% & Mon & \cellcolor{ForestGreen!5}9.55\% & \cellcolor{red!45}90.4\%  & \cellcolor{ForestGreen!10}19.5\%  & \cellcolor{red!40}80.5\% \\
		\cline{1-1} \cline{6-6}
		04:00 & \cellcolor{ForestGreen!3}5.88\%  & \cellcolor{red!47}94.1\%  & \cellcolor{ForestGreen!4}7.32\%  & \cellcolor{red!46}92.7\% & Tue & \cellcolor{ForestGreen!3}6.1\%  & \cellcolor{red!47}93.9\%  & \cellcolor{ForestGreen!9}17.1\%  & \cellcolor{red!41}82.9\% \\
		\cline{1-1} \cline{6-6}
		08:00 & \cellcolor{ForestGreen!2}3.92\%  & \cellcolor{red!48}96.1\%  & \cellcolor{ForestGreen!5}11\%  & \cellcolor{red!44}89\% & Wed & \cellcolor{ForestGreen!2}4.47\%  & \cellcolor{red!48}95.5\%  & \cellcolor{ForestGreen!5}11\%  & \cellcolor{red!44} 89\% \\
		\cline{1-1} \cline{6-6}
		12:00 & \cellcolor{ForestGreen!3}5.88\%  & \cellcolor{red!47}94.1\%  & \cellcolor{ForestGreen!7}14.6\%  & \cellcolor{red!44}85.4\% & Thu & \cellcolor{ForestGreen!4}7.52\%  & \cellcolor{red!46}92.5\%  & \cellcolor{ForestGreen!7}13.4\%  & \cellcolor{red!43}86.6\% \\
		\cline{1-1} \cline{6-6}
		16:00 & \cellcolor{ForestGreen!2}4.41\%  & \cellcolor{red!48}95.6\%  & \cellcolor{ForestGreen!7}13.4\%  & \cellcolor{red!43}86.6\% & Fri & \cellcolor{ForestGreen!4}7.11\%  & \cellcolor{red!46}92.9\%  & \cellcolor{ForestGreen!5}11\%  & \cellcolor{red!44}89\% \\
		\cline{1-1} \cline{6-6}
		20:00 & \cellcolor{ForestGreen!3}6.37\%  & \cellcolor{red!47}93.6\%  & \cellcolor{ForestGreen!9}17.1\%  & \cellcolor{red!41}82.9\% & Sat & \cellcolor{ForestGreen!4}7.32\%  & \cellcolor{red!46}92.7\%  & \cellcolor{ForestGreen!10}19.5\%  & \cellcolor{red!40}80.5\% \\		
\cline{6-6}
		 \multicolumn{5}{|l|}{\cellcolor[HTML]{EFEFEF}} & 		Sun & \cellcolor{ForestGreen!5}9.55\%  & \cellcolor{red!45}90.4\%  & \cellcolor{ForestGreen!10}20.7\%  & \cellcolor{red!40}79.3\% \\
		\hline				
	\end{tabular}
	\caption{Pair-wise tests of statistical significance of different times of the day and days of the week for each configuration. Values in the $p < 0.01$ columns represent percentage of pairs for which we found statistically significant differences.}
	\label{tab:other_factors}
\end{table}

As the resulting p-values are too numerous to fully enumerate here, and individually of little interest,
we only provide an aggregation in Table \ref{tab:other_factors}. For each time slot and
day, we provide the percentage of p-values over all configurations that were below and above the confidence level of $0.01$. Clearly, for each time slot and day, there are a
small number of pairs for which our tests showed significant differences. However, for the  majority of pairs, no statistically significant difference could be established. Further, we were not able to discover any systematics behind the significant pairs, leading us to believe that they represent statistical artifacts and outliers rather than relevant differences. Hence, we conclude that there is \emph{no support for the hypotheses \ref{hyp:time}-\ref{hyp:pday}} in our data.

\begin{figure}[h]
\centering
 \includegraphics[width=\linewidth]{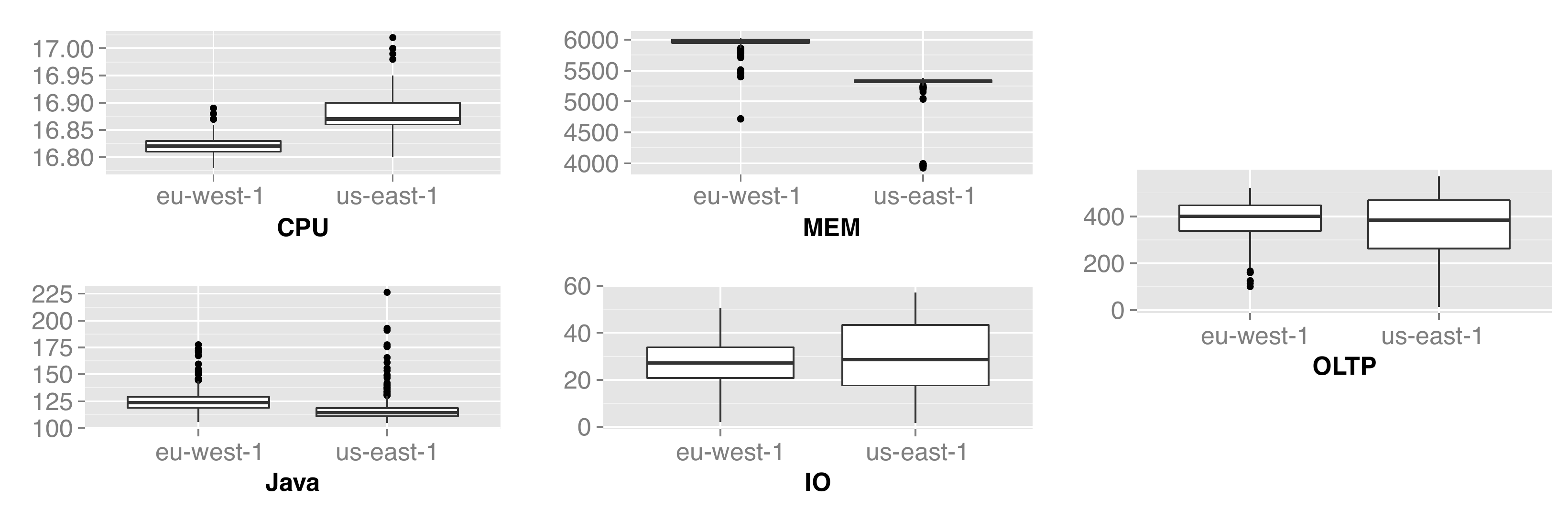}
 \caption{Distributions of benchmarked performance of \texttt{m3.large} EC2 instances in the \texttt{eu-west-1} and \texttt{us-east-1} regions.}
 \label{fig:regions}
\end{figure}

However, when comparing regions, results are different. In Figure~\ref{fig:regions}, we provide boxplots for all benchmarks
to compare the \texttt{eu-west-1} and \texttt{us-east-1} regions for the EC2 \texttt{m3.large} instance type. This time, visual analysis indicates
that there may indeed be a small but statistically significant difference in distribution between the regions, as projected in \ref{hyp:regions} and \ref{hyp:pregions}. More concretely (considering that for the CPU and Java benchmarks, lower is better) our data seems to indicate that \texttt{us-east-1} is typically worse-performing as well as more unpredictable (i.e., have higher standard deviation) than \texttt{eu-west-1}. Incidentally, this may explain why providers are, at the time of our research, typically pricing the European regions slightly higher than the North American regions.     

\begin{table}[ht]
	\scriptsize
	\centering
	\begin{tabular}{|l|l|l|c|c|c|c|c|}
		\hline
		\rowcolor[HTML]{EFEFEF} & & & \multicolumn{3}{c|}{\textbf{CPU-Bound}} & \multicolumn{2}{c|}{\textbf{IO-Bound}}\\
\hline
		\rowcolor[HTML]{EFEFEF} & \textbf{Type} & \textbf{Hyp.}  & \textbf{CPU} & \textbf{MEM} & \textbf{Java} & \textbf{IO} & \textbf{OLTP} \\ 
		\hline
		\multirow{6}{*}{\rotatebox[origin=c]{90}{EC2}} & \multirow{2}{*}{t1.micro} & H3.5 & \cellcolor{white!1}\cellcolor{white!1}* & \cellcolor{white!1}*  & \cellcolor{white!1}* & \cellcolor{white!1}* & \cellcolor{red!1} 0.016 \\
		\cline{3-3}
		& & H3.6 & \cellcolor{white!1}* & \cellcolor{white!1}* & \cellcolor{white!1}* & \cellcolor{white!1}* & \cellcolor{red!22} 0.221  \\
		\cline{2-8}
		& \multirow{2}{*}{m1.small} & H3.5 & \cellcolor{white!1}* & \cellcolor{white!1}*  & \cellcolor{white!1}* & \cellcolor{white!1}* & \cellcolor{white!1} * \\
		\cline{3-3}
		& & H3.6 & \cellcolor{white!1}* & \cellcolor{white!1}* & \cellcolor{white!1}* & \cellcolor{white!1}* & \cellcolor{white!1} *  \\
		\cline{2-8}
		& \multirow{2}{*}{m3.large} & H3.5 & \cellcolor{white!1}* & \cellcolor{white!1}*  & \cellcolor{white!1}* & \cellcolor{red!3}0.03 & \cellcolor{red!6} 0.065 \\
		\cline{3-3}
		& & H3.6 & \cellcolor{white!1}* & \cellcolor{red!19} 0.192 & \cellcolor{red!20} 0.202 & \cellcolor{white!1}* & \cellcolor{white!1}*  \\
		
		\hline
		
		\multirow{6}{*}{\rotatebox[origin=c]{90}{GCE}} & \multirow{2}{*}{f1-micro} 
		& H3.5 & \cellcolor{white!1}* & \cellcolor[HTML]{EFEFEF} & \cellcolor{red!67} 0.670  & \cellcolor{white!1}* & \cellcolor[HTML]{EFEFEF} \\
		\cline{3-3}
		& & H3.6 & \cellcolor{red!25} 0.251 & \cellcolor[HTML]{EFEFEF} & \cellcolor{red!5} 0.058 & \cellcolor{white!1}* &  \cellcolor[HTML]{EFEFEF} \\
		 \cline{2-4} \cline{6-7}
		& \multirow{2}{*}{n1-standard-1} & 
		H3.5 & \cellcolor{red!25} 0.253 & \cellcolor[HTML]{EFEFEF}  & \cellcolor{white!1}* & \cellcolor{white!1}* & \cellcolor[HTML]{EFEFEF} \\
		\cline{3-3}
		& & H3.6 & \cellcolor{white!1}*     & \cellcolor[HTML]{EFEFEF}  & \cellcolor{white!1}* & \cellcolor{white!1}* & \cellcolor[HTML]{EFEFEF}  \\
		\cline{2-4} \cline{6-7}
		& \multirow{2}{*}{n1-standard-2} & 
		H3.5 & \cellcolor{white!1}* & \cellcolor[HTML]{EFEFEF}  & \cellcolor{white!1}* & \cellcolor{white!1}* & \cellcolor[HTML]{EFEFEF} \\
		\cline{3-3}
		& & H3.6 & \cellcolor{white!1}* & \cellcolor[HTML]{EFEFEF}  & \cellcolor{white!1}* & \cellcolor{red!3} 0.034 & \cellcolor[HTML]{EFEFEF}  \\
		
		\hline		
		
		\multirow{6}{*}{\rotatebox[origin=c]{90}{Azure}} & \multirow{2}{*}{ExtraSmall} 
		& H3.5 & \cellcolor{white!1}* & \cellcolor[HTML]{EFEFEF} & \cellcolor{red!1} 0.017  & \cellcolor{white!1}* & \cellcolor[HTML]{EFEFEF} \\
		\cline{3-3}
		& & H3.6 & \cellcolor{red!22} 0.229 & \cellcolor[HTML]{EFEFEF} & \cellcolor{red!73}0.737 & \cellcolor{white!1}* &  \cellcolor[HTML]{EFEFEF} \\
		 \cline{2-4} \cline{6-7}
		& \multirow{2}{*}{Small} & 
		H3.5 & \cellcolor{white!1}* & \cellcolor[HTML]{EFEFEF}  & \cellcolor{white!1}* & \cellcolor{white!1}* & \cellcolor[HTML]{EFEFEF} \\
		\cline{3-3}
		& & H3.6 & \cellcolor{white!1}*     & \cellcolor[HTML]{EFEFEF}  & \cellcolor{white!1}* & \cellcolor{white!1}* & \cellcolor[HTML]{EFEFEF}  \\
		\cline{2-4} \cline{6-7}
		& \multirow{2}{*}{Medium} & 
		H3.5 & \cellcolor{red!34} 0.347 & \cellcolor[HTML]{EFEFEF}  & \cellcolor{white!1}* & \cellcolor{white!1}* & \cellcolor[HTML]{EFEFEF} \\
		\cline{3-3}
		& & H3.6 & \cellcolor{red!25} 0.259 & \cellcolor[HTML]{EFEFEF}  & \cellcolor{white!1}* & \cellcolor{red!41} 0.412 & \cellcolor[HTML]{EFEFEF}  \\
		
		\hline		
		
	\end{tabular}

	\caption{P-values for Wilcoxon rank-sum (H3.5) and Levene's variance (H3.6) test for comparing the performance in the \texttt{eu-west-1} and \texttt{us-east-1} regions. * codes statistically significant differences for a significance level of $p < 0.01$.}
	\label{tab:region}
\end{table}

We again validate our intuition using Wilcoxon signed-rank tests for \ref{hyp:regions}
and via Levene's variance test for \ref{hyp:pregions}. The assumed null-hypothesis for all tests was again
that there is \emph{no} difference in underlying distributions between regions.
We again tested for a confidence level of $0.01$. The resulting p-values for all configurations are
listed in Table~\ref{tab:region}. All statistically significant p-values (i.e., values $< 0.01$)
are coded as * for readability. 
We were able to find statistically significant differences
in 26 of 33 relevant configurations for \ref{hyp:regions},
and in 23 of 33 relevant configurations for \ref{hyp:pregions}. Hence, we
consider these hypotheses to be supported by our data.
The only cases for which we systematically could not find a significant difference is for the OLTP
benchmark in EC2, and for the CPU benchmark in Azure. Generally, in Azure, the difference in performance and predictabilty between
regions is much less pronounced than in EC2 and GCE.
Note that these tests do not make any statement about which region is actually \emph{preferable}
in which configuration. Using visual comparison of boxplots, we verified that indeed, when there is a difference, the analyzed European region is preferable to the North American region in the majority of cases.  



\subsubsection{Instance Type Selection}

In H4, we investigate the economic implications involved in selecting instance types. For \ref{hyp:diseconomies}, we examine whether the ratio of performance and costs tends to decrease with in increasing costs.
We consider on-demand, Linux-based cloud instances that are priced for a full hour. Further, we consider the prices that were valid
at the time we conducted our experiments, i.e., July/August 2014 for EC2 and GCE, and June/July 2015 for Azure and SL. The base hour prices are given in Table~\ref{tab:prices}. For these experiments, we are not interested in how instances of different providers compare, but rather in the comparison of cost-performance ratio (i.e., the ``value for money'') of different instance types in the same region of the same provider. Hence, as a first step, we normalize all benchmark results to the mean value of a small instance type (i.e., \texttt{m1.small} for EC2, \texttt{n1-standard-1} for GCE, \texttt{Small} for Azure, and the single-core machine ine SL) of the same provider and region, as in Equation \ref{eq:rel_performance}. Further, for the CPU and Java benchmark, we also invert the value, to achieve a consistent ``higher-is-better'' notion across all benchmarks.

\begin{equation}
\label{eq:rel_performance}
\forall c \in \mathcal{C} \; : \; \overline{m_c^*} = 
\begin{cases}
\frac{\overline{m_c}}{\overline{m_c}(small)}, \text{\scriptsize{for (MEM, IO, OLTP)}} \\[2ex]
\frac{\overline{m_c}(small)}{\overline{m_c}}, \text{\scriptsize{for (CPU, Java)}}
\end{cases}
\end{equation}

Similarly, we normalize all hourly prices as in Table~\ref{tab:prices} to the costs of a small instance of the same provider and region (Equation \ref{eq:rel_costs}).

\begin{equation}
\label{eq:rel_costs}
\forall c \in \mathcal{C} \; : \; p_c^* = \frac{p_c}{p_c(small)}
\end{equation}

Now we can define the normalized cost-performance ratio $cpr_c^*$ for every possible instance configuration $c \in \mathcal{C}$ (Equation \ref{eq:cost_performance}). Per definition, all small instances have a $cpr_c^*$ of exactly 1. For all other instance types, a $cpr_c^* < 1$ represents a cost-performance ratio worse than a small instance in the same region and provider, while a $cpr_c^* > 1$ represents better
cost efficiency. Note again that these values cannot be compared between providers, regions, or benchmarks, as they are normalized to different base values.

\begin{equation}
\label{eq:cost_performance}
\forall c \in \mathcal{C} \; : \; cpr_c^* = \frac{\overline{m_c^*}}{p_c^*}
\end{equation}

\begin{figure}[h]
	\centering
	\includegraphics[width=\linewidth]{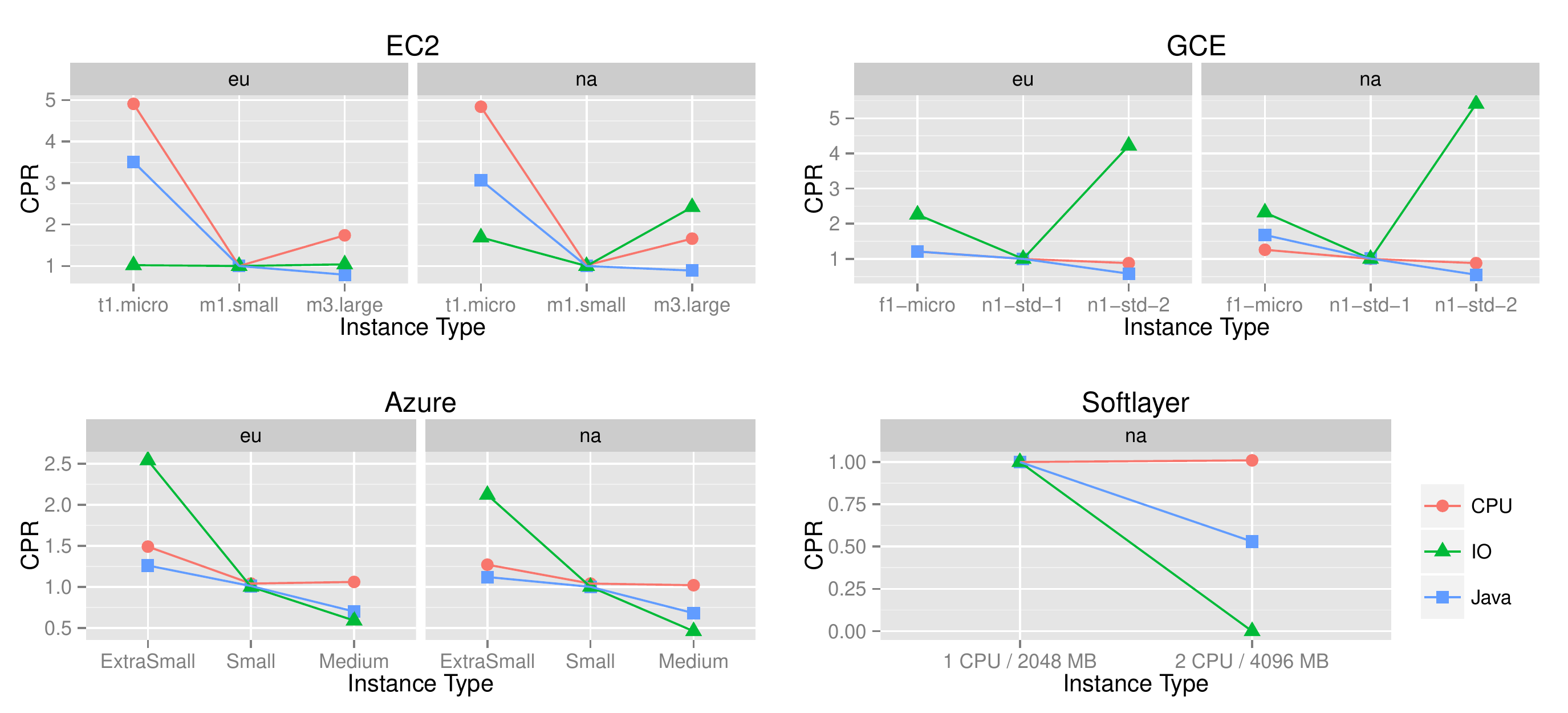}
	\caption{Cost-performance ratio \rev{for each} provider, region, instance type, and benchmark.}
	\label{fig:cost_performance_ratio}
\end{figure}

\begin{wraptable}{r}{0.4\textwidth}
\centering
\scriptsize
\begin{tabular}{|l|l|l|l|}
\hline
\rowcolor[HTML]{EFEFEF} &  & \textbf{Type} & \textbf{Costs} \\
\hline
\multirow{8}{*}{\rotatebox[origin=c]{90}{EC2}} & \multirow{5}{*}{\rotatebox[origin=c]{90}{eu}}  & t1.micro & 0.020\$ \\
\cline{3-3}
 &  & m1.small & 0.047\$ \\
 \cline{3-3} 
&  & m3.large & 0.154\$ \\
\cline{3-3} 
&  & c3.large & 0.120\$\\
\cline{3-3} 
&  & i2.xlarge & 0.938\$ \\
\cline{2-3} 
& \multirow{3}{*}{\rotatebox[origin=c]{90}{na}}  & t1.micro & 0.020\$ \\
\cline{3-3}
 &  & m1.small & 0.044\$\\
 \cline{3-3}
&  & m3.large & 0.140\$\\
\hline 
\multirow{6}{*}{\rotatebox[origin=c]{90}{GCE}} & \multirow{3}{*}{\rotatebox[origin=c]{90}{eu}} & f1-micro & 0.013\$\\
\cline{3-3}
 &  & n1-standard-1 & 0.069\$\\
 \cline{3-3}
 &  & n1-standard-2 & 0.138\$\\
 \cline{2-3}
 & \multirow{3}{*}{\rotatebox[origin=c]{90}{na}} & f1-micro & 0.012\$\\
 \cline{3-3}
 &  & n1-standard-1 & 0.063\$\\
 \cline{3-3}
 &  & n1-standard-2 & 0.126\$\\
 \hline
\multirow{6}{*}{\rotatebox[origin=c]{90}{Azure}} & \multirow{3}{*}{\rotatebox[origin=c]{90}{eu}} & ExtraSmall & 0.018\$\\
\cline{3-3}
 &  & Small & 0.051\$ \\
 \cline{3-3}
 &  & Medium & 0.102\$\\
 \cline{2-3}
 & \multirow{3}{*}{\rotatebox[origin=c]{90}{na}} & ExtraSmall & 0.018\$\\
 \cline{3-3}
 &  & Small & 0.044\$\\
 \cline{3-3}
 &  & Medium & 0.088\$ \\
 \hline 
\multirow{2}{*}{\rotatebox[origin=c]{90}{SL}} & \multirow{2}{*}{\rotatebox[origin=c]{90}{na}} & 1 CPU / 2048 MB & 0.053\$\\
\cline{3-3}
 &  & 2 CPUs / 4096 MB & 0.105\$ \\
 \hline  
\end{tabular}
\caption{Hourly prices of all configurations.}
\label{tab:prices}
\end{wraptable}
In Figure~\ref{fig:cost_performance_ratio}, we depict $cpr_c^*$ for all micro, small and large instance types in our study for the CPU, IO and Java benchmarks. \ref{hyp:diseconomies} claims that larger (i.e., more expensive) instance types are generally less cost-efficient. Our data only supports this hypothesis for Azure and Softlayer. For EC2 and GCE, \texttt{m1.small} and \texttt{n1-standard-1} instances
are often the least cost-efficient instance type in our study. Generally, which instance types are the most cost-efficient choice
is highly provider and use case specific. Hence, we conclude that \ref{hyp:diseconomies} is not generally supported by our experiments. Contrary, \ref{hyp:stability} claims that the performance of larger instance types is more predictable than of cheaper instance types. The data already presented in Table~\ref{tab:inter} supports this hypothesis for CPU-bound benchmarks in EC2 and GCE, but there is no
strong support for this hypothesis for IO-bound benchmarks or the other two providers. Similarly to \ref{hyp:diseconomies}, we conclude
that no clear, provider- and benchmark-independent statement can be made in this regard.

\begin{wrapfigure}{r}{0.6\textwidth}
	\centering
	\includegraphics[width=\linewidth]{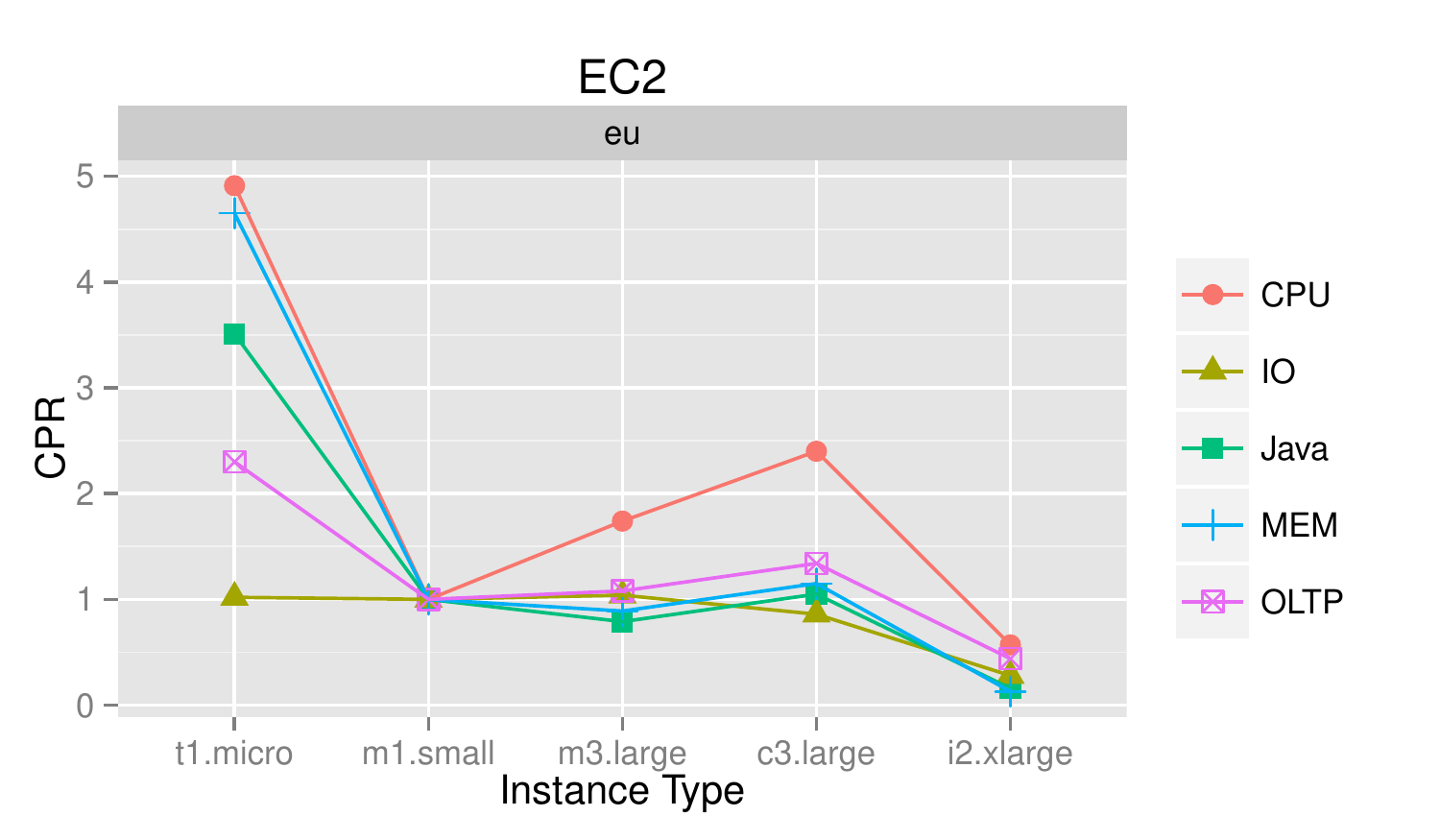}
	\caption{Cost-performance ratio for all benchmarks in the European region of EC2.}
	\label{fig:cost_performance_ratio_h43}
\end{wrapfigure}

Following \ref{hyp:special}, we expected specialized instance types to be more cost-efficient in benchmarks that relate to their ``speciality'', such as IO-bound benchmarks for IO-optimized instance types.
We have collected data related to specialized instance types only for EC2 and
the European region. This data is depicted in Figure~\ref{fig:cost_performance_ratio_h43}.
For the CPU-optimized instance type \texttt{c3.large}, we indeed observe a slightly better cost-performance ratio for CPU-bound benchmarks than for the \texttt{m1.small}
and \texttt{m3.large} instance types. However, for the IO-optimized \texttt{i2.xlarge} instance type, even the cost-performance ratio
in the IO benchmark is relatively unfavorable in comparison to all other instance types.
This is largely due to the very high hourly costs of this instance type, which is not fully made up for with increased performance. Concluding, cloud users should not fall into the trap
of assuming that specialized instance types are always the best choice for a use case that requires substantial IO or CPU. The often over-proportionally higher price for specialized instance types should always be taken into account when
selecting between those and general-purpose instance types. However, users need to keep in mind that some options \rev{may} only available for specialized instance types.

\section{Implications}

We now discuss the main
outcomes and implications of our study.  

\textbf{\emph{Substantial differences between providers.}}
For {\ref{hyp:inter}-\ref{hyp:lifetime_bursting}}, which relate to the
fundamental properties of performance predictability, we found EC2 and Azure to largely
behave as we expected after analyzing existing research.
GCE and SL, on the other hand, were much more stable and predictable
than what we anticipated based on the literature. This leads us
to believe that currently there is too much focus on benchmarking EC2, and that
the research community needs to validate to what extent existing
results are generalizable to other cloud providers.
 We argue that a wide-ranging survey
across a large number of IaaS providers is required to reality-check some of the
cloud computing communities' standing assumptions. Our work serves as a first
step into this direction.

\textbf{\emph{Lack of hardware heterogeneity.}}
One such assumption is the concept of hardware heterogeneity, on which a large
body of scientific research rests (e.g.,~\cite{ou:13,farley:12}).
Based on our experiments, we conclude that  hardware heterogeneity is today much
less of a factor than what previous studies have reported. We can only speculate
as to the reasons of this discrepancy. However, it is likely that cloud providers
have simply reacted to commercial pressure to deliver more predictable ``value-for-money''
in comparison to the early years of the cloud hype.
Currently, hardware heterogeneity in EC2 is really only important
for micro and small instance types (and even for those only in the North American region).
All more expensive instance types (which are arguably also more relevant in
industrial practice) are currently served with a fixed CPU model in EC2.
In our study,
only Azure still fully embraces the idea of hardware heterogeneity for all instance types.

\textbf{\emph{Multi-tenancy is indeed important, but not equally so for all providers.}}
As expected, multi-tenancy is a large factor in cloud performance predictability, but
the extent of this varies substantially between providers. For the more mainstream cloud providers
EC2 and Azure, all IO-bound benchmarks are highly unpredictable, while the more niche players
GCE and SL actually perform rather predictably even for IO-bound benchmarks.
We speculate that this is due to higher infrastructure
utilization in the former cloud providers. In any case, our results show that the research
community needs to more carefully cross-check
results across multiple clouds, as research results
can vary substantially between providers even for long-standing assumptions such as
the importance of multi-tenancy.

\textbf{\emph{Regions matter, time and day do not.}}
For {\ref{hyp:time}-\ref{hyp:pregions}}, which capture the temporal and
geographical factors that
influence performance and predictability, we found little empirical evidence that either
the day of the week or the time of the day have any real, measurable impact on
performance in any provider. However, different regions perform differently in all providers
for most benchmarks, but the difference is considerably less pronounced in GCE
and Azure than in EC2.

\textbf{\emph{Selecting the right instance type is non-trivial.}}
Our results relating to the cost-performance ratio of cloud instances shows that
there are no reliable ``rules of thumb'' for selecting the right instance types.
Different providers employ different pricing strategies, and it is rarely clear
without explicit benchmarking which instance type of a provider is particularly
cost-efficient for which use case. This is also true for specialized instance
types, which are not necessarily more cost-efficient in their specialization than
general-purpose types.
Arguably, the selling point of, for example, IO-optimized instances is rather the
high ceiling of performance and the possibility to use special configuration options.
Careful evaluation and benchmarking is still essential to get the best performance and predictability
per US dollar spent. 


%

\section{Threats to Validity}

As with any empirical research, there are threats and limitations to our study, which we
discuss in the following.

\textbf{Construct Validity.} Both during literature review and
experimentation, some design decisions had to be made. Most importantly,
our choice of method for literature search (seeding the search with well-known
standard publications and then following the citation graph) has the threat of
missing relevant ``unconnected'' papers (e.g., from different research communities). We
have mitigated this risk by verifing for 10 arbitrary relevant papers
(found by searching the
ACM Digital Library\footnote{\url{http://dl.acm.org}}) that each they were indeed
contained in our study set. Another design decision was which benchmarks and
configurations to select, and how to configure and parameterize each benchmark.
Our approach here was primarily to use common ``out-of-the-box'' tools and
configurations as far as possible, to prevent us from biasing the results by
unbalanced optimization.

\textbf{Internal Validity.} We collected an extensive amount of data to
validate our hypotheses. However, we did so during a relatively short period of time.
This increases the threat that external factors influenced our results (e.g.,
exceptionally high load during summer in EC2).
More long-term research is required to control for this threat, which was out
of scope of our present work. 

\textbf{External Validity.} We necessarily had to select a subset from the
vast space of available cloud providers, instance types, regions, and
configuration options. This leads to the question to what extent our results are
generalizable to other providers and configurations. Indeed, one of our
central outcomes is that benchmarking results vary substantially between different
cloud providers. However, our choice of providers included the two current market leaders
(EC2 and Azure), one up-and-coming provider (GCE), and a niche player (SL). Hence,
we argue that our coverage of the IaaS market is sufficient to support our conclusions.


\section{Conclusions and Outlook}
\label{sec:conclusions}

We have conducted a structured analysis of the fundamental principles
of performance variation and predictability in public
IaaS providers. We have analyzed the current state of research, and
formulated 15 hypotheses based on previously published, peer-reviewed
literature. Further, we have validated these hypotheses based on real-life
data collected from Amazon's EC2, Google's GCE, Microsoft's Azure, and
IBM's Softlayer clouds.
Analysis of
this data showed that there are substantial differences in the performance
of different providers, and that practitioners should not assume that research
on EC2 is necessarily applicable to other providers. Further,
we have observed that hardware heterogeneity is less of a practical factor
than what earlier research has reported. Multi-tenancy is indeed important,
but not to the same extent for all providers. These results indicate that cloud performance
is indeed a ``moving target'', and that the scientific community is required to
periodically re-validate its understanding of the subject.

We have not been able to establish conclusive causality between different days of
the week, or times of the day, and observed performance. However, as expected, the
region has a statistically significant impact on performance and
predictability on both providers. Finally, we have seen that it is not feasible to
establish hard and fast rules for cost-efficient cloud instance selection. Users are required to
benchmark instance types for their specific use cases.

One key issue that our research did not address is
the longer-term performance stability of cloud instances. Our
own work only studied performance stability within cloud instances in the time
frame of three days, and we are not aware of any other publication that
systematically benchmarked instances for a longer period.
Doing so
would help practitioners get a feel for how frequently they need to re-evaluate
the performance of the instances they are using. More generally, our results 
call for a more longitudinal study, which tracks the performance of IaaS providers
over multiple months or even years, to decide whether the observations reported in
this paper (as well as in previous work) are subject to temporal (e.g., seasonal)
changes.

\section*{Companion Website}
We maintain an online companion\footnote{\url{http://wp.ifi.uzh.ch/leitner/?p=588}}, which
contains
the used benchmark code,
and all data used in our study. Further, we will provide
additional plots as well as scripts used for analysis
and data cleaning. 


\bibliographystyle{ACM-Reference-Format-Journals}
\bibliography{bibtex}

\end{document}